\documentclass[aps,prd,nofootinbib,superscriptaddress,11pt]{revtex4-1}

\usepackage{graphicx}
\usepackage{color}
\usepackage{bm}
\usepackage{ulem}
\usepackage{setspace}
\usepackage{amsfonts}
\usepackage{multirow}

\newcommand\deleted[1]{}

\makeatletter
\newtoks\ignoretoks
\newcommand\question{\afterassignment\egroup\question@\bgroup}
\newcommand\question@{\ignoretoks=}
\let\@orig@footnotetext=\@footnotetext
\renewcommand\@footnotetext[1]{\@orig@footnotetext{%
                                 \relax\baselineskip0pt\relax
                                 #1}}
\makeatother

\usepackage{xr}
\newcommand{\dtrip}{D_{\mathrm{triplet}}}
\newcommand{\drf}{D_{\mathrm{RF}}}

\begin{document}

\title{On the universal structure of human lexical semantics}

\author{Hyejin Youn}
\affiliation{Institute for New Economic Thinking at the Oxford Martin School, Oxford, OX2 6ED, UK}
\affiliation{Mathematical Institute, University of Oxford, Oxford, OX2 6GG, UK}
\affiliation{Santa Fe Institute, 1399 Hyde Park Road, Santa Fe, NM 87501, USA}
\author{Logan Sutton}
\affiliation{Department of Linguistics, University of New Mexico, Albuquerque, NM 87131, USA}
\author{Eric Smith}
\affiliation{Santa Fe Institute, 1399 Hyde Park Road, Santa Fe, NM 87501, USA}
\author{Cristopher Moore}
\affiliation{Santa Fe Institute, 1399 Hyde Park Road, Santa Fe, NM 87501, USA}
\author{Jon F. Wilkins}
\affiliation{Santa Fe Institute, 1399 Hyde Park Road, Santa Fe, NM 87501, USA}
\affiliation{Ronin Institute, Montclair, NJ 07043}
\author{Ian Maddieson}
\affiliation{Department of Linguistics, University of New Mexico, Albuquerque, NM 87131, USA}
\author{William Croft}
\affiliation{Department of Linguistics, University of New Mexico, Albuquerque, NM 87131, USA}
\author{Tanmoy Bhattacharya}
\affiliation{Santa Fe Institute, 1399 Hyde Park Road, Santa Fe, NM 87501, USA}
\affiliation{MS B285, Grp T-2, Los Alamos National Laboratory, Los Alamos, NM 87545, USA.}


\date{\today}

\begin{abstract}
{How universal is human conceptual structure? The way concepts are
organized in the human brain may reflect distinct features of
cultural, historical, and environmental background in addition to
properties universal to human cognition. Semantics, or meaning 
expressed through language, provides direct access to the 
underlying conceptual structure, but meaning is notoriously difficult 
to measure, let alone parameterize.  Here we provide an empirical 
measure of semantic proximity between concepts using cross-linguistic 
dictionaries.
Across languages carefully selected from a phylogenetically and 
geographically stratified sample of genera, translations of words 
reveal cases where a particular language uses a single polysemous word 
to express concepts represented by distinct words in another.  We use 
the frequency of polysemies linking two concepts as a measure of their 
semantic proximity, and represent the pattern of such linkages by a 
weighted network.  This network is highly uneven and fragmented: certain 
concepts are far more prone to polysemy than others, and there emerge 
naturally interpretable clusters loosely connected to each 
other.  Statistical analysis shows such structural properties are 
consistent across different language groups,
largely independent of geography, environment, 
and literacy. It is therefore possible to conclude the conceptual structure 
connecting basic vocabulary studied is primarily due to universal features 
of human cognition and language use.  }
\end{abstract}

\maketitle 

The space of concepts expressible in any language is vast.  This space 
is covered by individual words representing semantically tight 
neighborhoods of salient concepts.  There has been much debate about whether 
semantic similarity of concepts is shared across 
languages \cite{Whorf1956,Fodor1975, Wierzbicka1996,Lucy1992,Levinson2003,
ChoiBowerman1991, Majid2008,Croft2010}.
On the one hand, all human beings belong to a single species characterized 
by, among other things, a shared set of cognitive abilities. On the other 
hand, the 6000 or so extant human languages spoken by different societies 
in different environments across the globe are extremely 
diverse~\cite{EvansLevinson2009,Comrie1989,Croft2003} and may reflect 
accidents of history as well as adaptations to local environments.
Most psychological experiments about this question have been conducted
on members of ``WEIRD'' (Western, Educated, Industrial, Rich,
Democratic) societies, yet there is reason to question whether the 
results of such research are
valid across all types of
societies \cite{Henrich2010}. Thus, the question of the degree to which
conceptual structures expressed in language are due to universal
properties of human cognition, the particulars of cultural history, or
the environment inhabited by a society, remains unresolved.

The search for an answer to this question has been hampered by a major
methodological difficulty. Linguistic meaning is an abstract construct
that needs to be inferred indirectly from observations, and hence is
extremely difficult to measure; this is even more apparent in the
field of lexical semantics. Meaning thus contrasts both with
phonetics, in which instrumental measurement of physical properties of
articulation and acoustics is relatively straightforward, and with grammatical structure, 
for which there is general agreement on a number of basic units of
analysis \cite{Shopen2007}. Much lexical semantic analysis relies on linguists'
introspection, and the multifaceted dimensions of meaning currently
lack a formal characterization.  To address our primary question, it
is necessary to develop an empirical method to characterize the space
of lexical meanings.

We arrive at such a measure by noting that translations uncover the
alternate ways that languages partition meanings into words. Many words 
have more than one meaning, or sense, to the extent that word senses can 
be individuated \cite{CroftCruse2004}.  Words gain meanings when their use
is extended by speakers to similar meanings; words lose meanings when 
another word is extended to the first word's meaning, and the first word 
is replaced in that meaning.  
To the extent that words in transition across similar, or possibly contiguous, 
meanings account for the polysemy (multiple meanings of a single word form)
revealed in cross-language translations, the frequency of polysemies found across an 
unbiased sample of languages can provide a measure of semantic 
similarity among word meanings.  
The unbiased sample of languages is carefully chosen 
in a phylogenetically and geographically stratified way, according to 
the methods of typology and universals research \cite{Comrie1989,Croft2003}.  
This large, diverse sample of languages allows us to avoid the 
pitfalls of research based solely on ``WEIRD'' societies and to separate 
contributions to the empirically attested patterns in the linguistic data, 
arising from universal language cognition versus those from artifacts of the
speaker-groups' history or way of life.


There have been several cross-linguistic surveys of lexical polysemy, 
and its potential for understanding semantic shift \cite{Koptjevskaja2012}, 
in the domains such as body parts \cite{Brown1976,WitkowskiBrown1978}, 
cardinal directions \cite{Brown1983}, perception verbs \cite{Viberg1983}, 
concepts associated with fire \cite{Evans1992}, and
color metaphors \cite{Derrig1978}.  
We add a new dimension to the existing body of research by providing
a comprehensive mathematical method using a systematically 
stratified global sample of languages to measure degrees of similarity. 
Our cross-linguistic study takes the Swadesh lists as basic concepts 
\cite{Swadesh1952} as  
most languages have words for them. Among those concepts, we chose 22 
meanings associated with two domains: celestial objects 
(e.g. \texttt{SUN, MOON, STAR}) 
and landscape objects (e.g. \texttt{FIRE, WATER, MOUNTAIN, DUST}).  For each 
word expressing one of these meanings, we examined what other concepts were 
also expressed by the word. 
Since the semantic structures of these two domains are very likely 
to be influenced by the physical environment that human societies 
inhabit, any claim of universality of lexical semantics needs to be 
demonstrated here.


\begin{figure*}
  \begin{center} 
 \includegraphics[width=.6\textwidth]{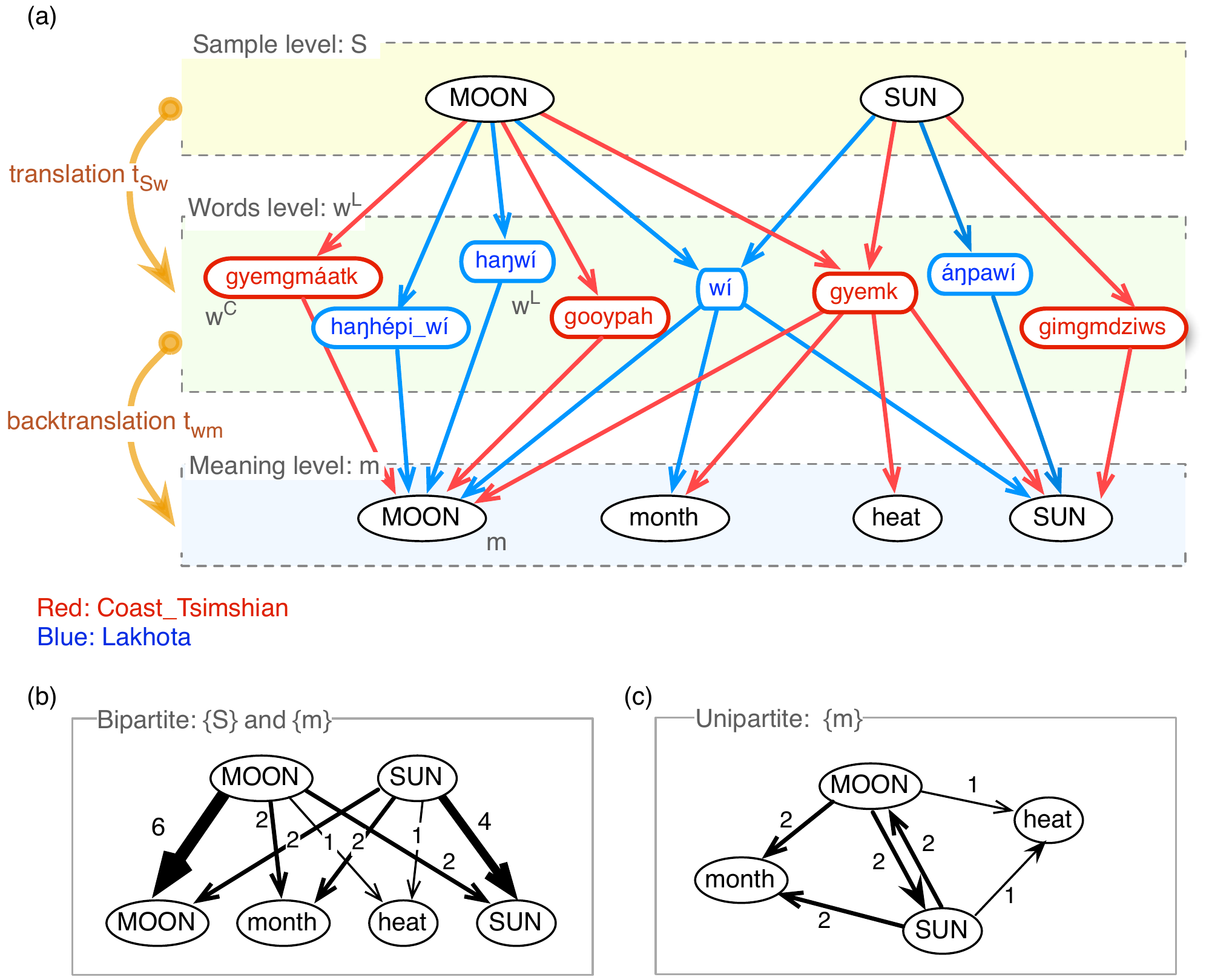}
  \caption{Schematic figure of the construction of network representations.  
		(a) Tripartite polysemy network constructed through translation (links 
		from the first to the second layer) and back-translation (links from 
		the second to the third layer) for the cases of {\ttfamily MOON} and 
		{\ttfamily SUN} in two American languages: Coast Tsimshian (red links) 
		and Lakhota (blue links).  
		(b) Directed bipartite graph of two languages grouped, projected from 
		the tripartite graph above 
		by aggregating links in the second layer.  (c) Directed and weighted 
		unipartite graph, projected from the bipartite graph by identifying and merging 
		the same Swadesh words ({\ttfamily MOON} and {\ttfamily SUN} in this case).  
%
  \label{fig:schematic} }
  \end{center}
\end{figure*}  

\section*{Results}
We represent word-meaning and meaning-meaning relations uncovered by 
translation dictionaries between each language in the unbiased sample 
and major modern European languages by constructing a network structure. 
Two meanings (represented by a set of English words) are linked if they are 
translated from one to another and then back, and the link is weighted by 
the number of paths of the translation, or the number of words that represent
both meanings (see Methods for detail). 
Figure \ref{fig:schematic} illustrates the construction in the case of 
two languages, Lakhota (primarily spoken in North and South Dakota) 
and Coast Tsimshian (mostly spoken in northwestern British Columbia and 
southeastern Alaska).
Translation of {\ttfamily SUN} in Lakhota results {\it w\'{i}}
and {\it \'a$\eta$paw\'{i}}. While the later picks up no other meaning,
{\it w\'{i}} is a polysemy that possesses additional meanings of
{\ttfamily MOON} and {\ttfamily month}, hence they are linked to {\ttfamily SUN}. 
Such polysemy is also observed in Coast Tsimshian where {\it gyemk}, 
translated from {\ttfamily SUN}, covers additional meanings including, 
thus additionally linking to, {\ttfamily heat}. 

Each language has its own way of partitioning meanings by words, captured in 
a semantic network of the language. It is conceivable, however, that a group 
of languages bear structural resemblance perhaps because the speakers share 
historical or environmental features. A link between {\ttfamily SUN} and {\ttfamily MOON}, 
for example, reoccurs in both languages, but does not appear in many 
other languages.  {\ttfamily SUN} is instead linked to {\ttfamily divinity} 
and {\ttfamily time} in Japanese, and to {\ttfamily thirst} and 
{\ttfamily DAY/DAYTIME} in !X\'o\~o.  The question then is the degree to 
which the observed polysemy patterns are general or sensitive to the 
environment inhabited by the speech community, phylogenetic 
history of the languages, and intrinsic linguistic factors such as literary 
tradition.  We test such question by grouping the individual networks in 
a number of ways according to properties of their corresponding languages.  
We first analyze the networks of the entire languages, and then of sub-groups. 

\begin{figure*}
  \begin{center} 
  \centerline{\includegraphics[width=.9\textwidth]{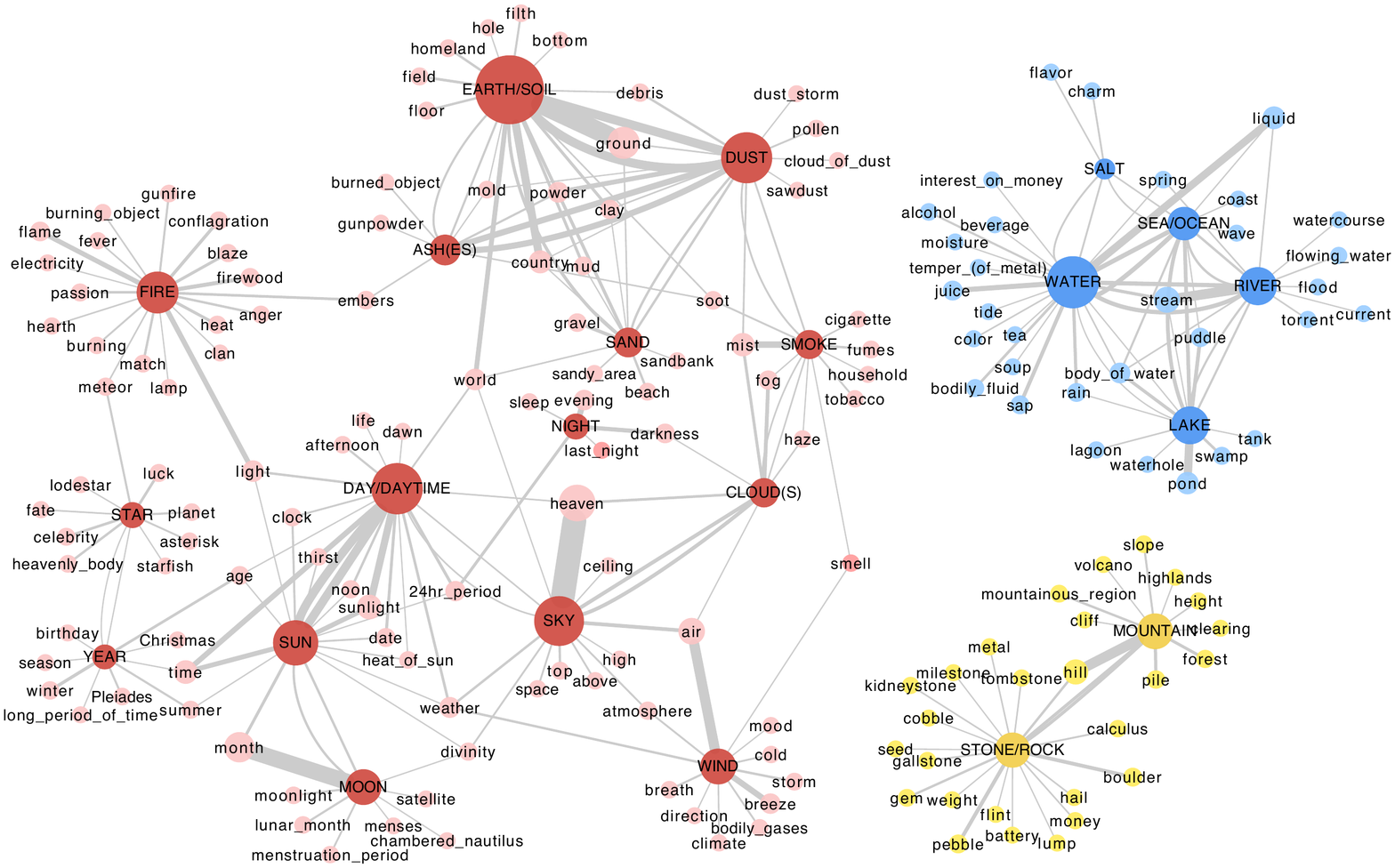}}
  \caption{ Connectance graph of concepts. Concepts are connected 
    through polysemous words that cover the concepts.  
		Swadesh entries are capitalized. 
		Links whose weights are more than two are presented, and direction is omitted 
		for simplicity. 
		The size of a node and the width of a link to another node 
		are proportional to the number of polysemies associated with 
		the concept, and with the two connected concepts, respectively. 
		The thick link from {\ttfamily SKY} to {\ttfamily heaven} denotes 
		the large number of polysemies across languages.  Three distinct 
		clusters are identified and coloured by red, blue, and yellow, 
		which may imply a coherent set of relationship among concepts that 
		possibly reflects human cognitive conceptualization of these semantic 
		domains.} \label{fig:connectance_graph} 
  \end{center}
\end{figure*}

In Fig.~\ref{fig:connectance_graph}, we present the network of the entire 
languages exhibiting the broad topological structure of polysemies observed 
in our data.
It reveals three almost-disconnected clusters, groups of concepts that are 
indeed more prone to polysemy within, that are associated with a natural 
semantic interpretation.  The semantically most uniform cluster, colored in 
blue, includes concepts related to water. 
A second, smaller cluster, colored in yellow, associates solid natural
substances (centered around {\ttfamily STONE/ROCK}) with their
topographic manifestation ({\ttfamily MOUNTAIN}).  The third cluster,
in red, is more loosely connected, bridging a terrestrial cluster and
a celestial cluster, including less tangible substances such as
{\ttfamily WIND}, {\ttfamily SKY}, and {\ttfamily FIRE}, and salient time
intervals such as {\ttfamily DAY} and {\ttfamily YEAR}.
In keeping with many traditional oppositions between {\ttfamily EARTH}
and {\ttfamily SKY/heaven}, or {\ttfamily darkness}, and {\ttfamily
  light}, the celestial, and terrestrial components form two
sub-clusters connected most strongly through {\ttfamily CLOUD}, 
which shares properties of both.  
The result reveals a coherent set of relationships among concepts
that possibly reflects human cognitive conceptualization of these
semantic domains \cite{Croft2003,Croft2010,Vygotsky:thought_lang:02}.


We test whether these relationships are universal rather than particular 
to properties of linguistic groups such as physical environment that 
human societies inhabit. 
We first categorized languages by nonlinguistic variables such as geography,
topography, climate, and the existence or nonexistence of a literary
tradition (Table~\ref{tab:groups} in Appendix) and constructed a network 
for each group.  A spectral algorithm then clusters Swadesh entries into
a hierarchical structure or dendrogram for each language group.
Using standard metrics on trees \cite{Critchlow:96,Dobson:75,Robinson-Foulds}, 
we find that the dendrograms of language groups are much closer 
to each other than to dendrograms of randomly permuted leaves:
thus the hypothesis that languages of different subgroups share 
no semantic structure in common is rejected 
($p < 0.05$, see Methods)---{\ttfamily SEA/OCEAN} and 
{\ttfamily SALT} are, for example, more related than either is to 
{\ttfamily SUN} in every group we tried.  
In addition, 
the distances between dendrograms of language groups are statistically 
indistinguishable from the distances between bootstrapped languages 
($p<0.04$). 
Figure 3 shows a summary of the statistical tests of 11 different groups.
Thus our data analyses provide consistent evidences that all languages 
share semantic structure, the way concepts are clustered in Fig. 2, with 
no significant influence from environmental or cultural factors.

\begin{figure*}[ht]
	\begin{center}
	\centerline{\includegraphics[width=.9\textwidth]{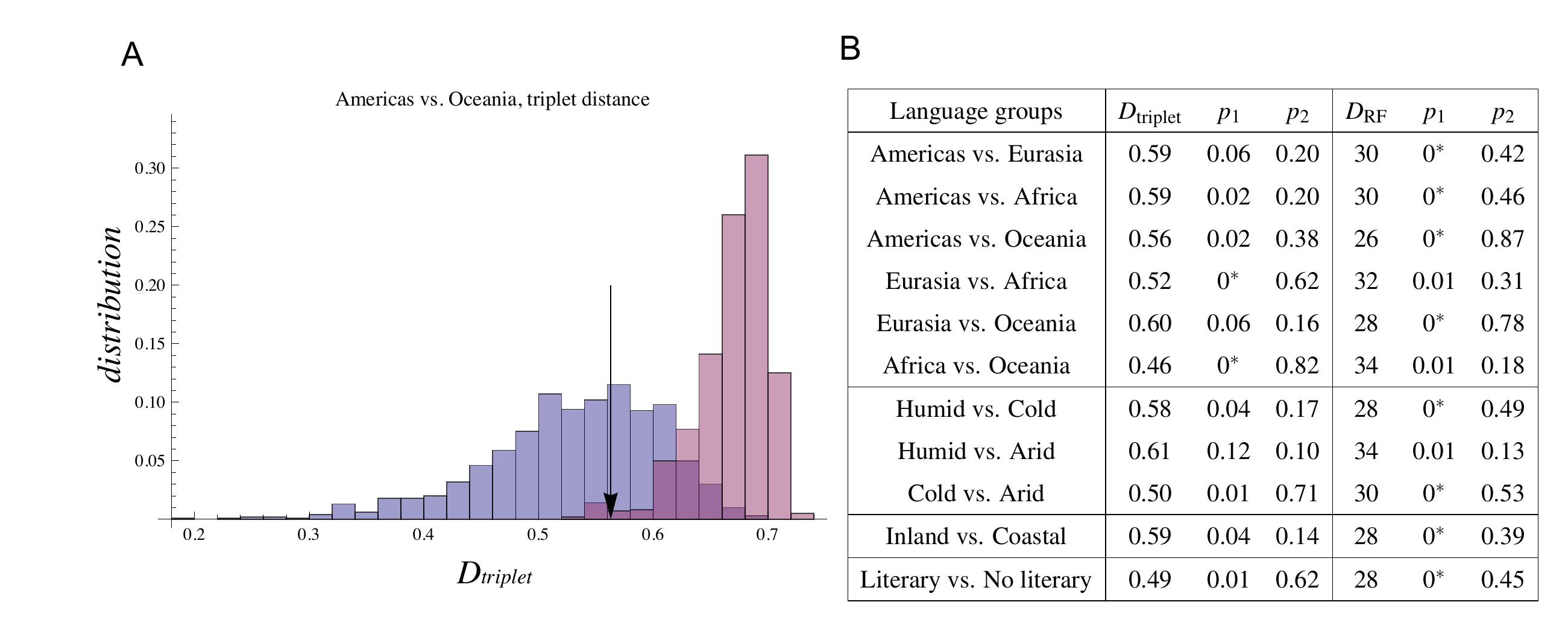}}
	\caption{ (A) An illustration of our bootstrap experiments.  The triplet 
	distance $\dtrip$ between the dendrograms of the Americas and Oceania is 
	$0.56$ (arrow).  This number sits at the very low end of the distribution 
	of distances when leaves are randomly permuted (the red shaded profile on 
	the right), but it is well within the distribution that we obtain by 
	randomly re-sampling from the set of languages (the blue shaded profile on 
	the left).  This gives strong evidence that each pair of subgroups share 
	an underlying semantic network, and that the differences between them are 
	no larger than would result from random sampling. (B) Comparing distances 
	(the triplet $\dtrip$ and the Robinson-Foulds $\drf$) among dendrograms 
	of subgroups and two types of bootstrap experiments: permuting leaves of 
	the dendrogram and replacing the subgroups in question with bootstrapped 
	samples of the same sizes. $p_1$-values for the former bootstrap 
	($p_2$-values for the latter) are the fraction of $1000$ bootstrap samples 
	whose distances are smaller (larger) than the observed distance.  
	In either case $0^\ast$ denotes a value below $0.001$, i.e., no bootstrap 
	sample satisfied the condition.} \label{fig:bootstrap}
\end{center}
\end{figure*}

\begin{figure}
  \centerline{\includegraphics[width=.7\textwidth]{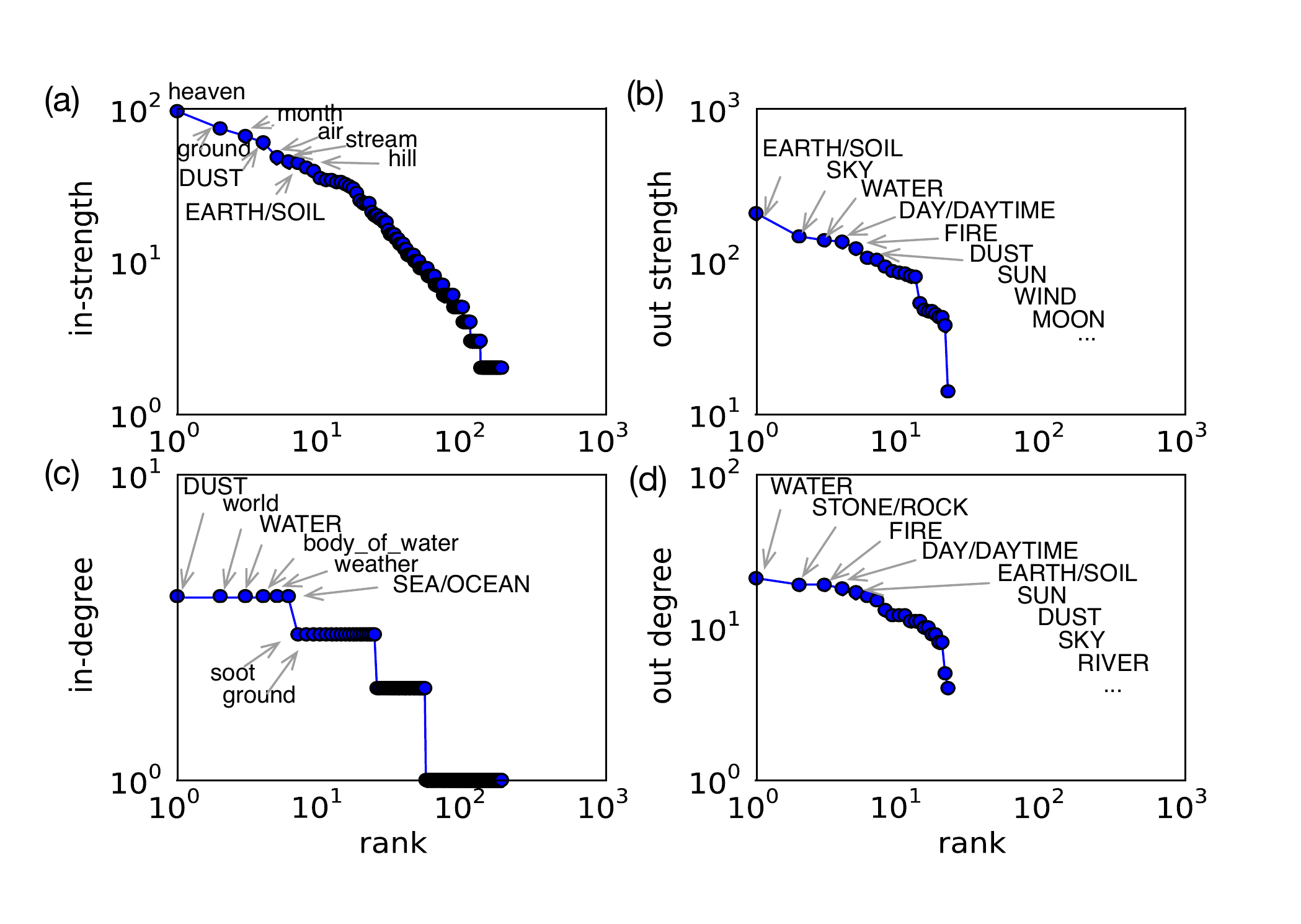}}
  \caption{Rank plot of concepts in descending order of their strengths 
		(summation of weighted links) and degrees (summation of unweighted links) 
		shown in Fig.~\ref{fig:connectance_graph}.  Entries from the initial 
		Swadesh list are distinguished with capital letters.  (a) in-strengths 
		of concepts: sum of weighted links to a node.  (b) out-strengths of
		Swadesh entries: sum of weighted links from a Swadesh entry. (c) degree 
		of the concepts: sum of unweighted links to a node  (d) degree of Swadesh 
		entries: sum of unweighted links to a node.  A node strength in this 
		context indicates the total number of polysemies associated with the concept 
		in 81 languages while a node degree means the number of other concepts 
		associated with the node regardless of the number of synonymous 
		polysemies associated with it.  {\ttfamily heaven}, for example, has the 
		largest number of polysemies, but most of them are with {\ttfamily SUN}, 
		so that its degree is only three.  \label{fig:word_degree_rank} }
\end{figure}

Another structural feature apparent in Fig.~\ref{fig:connectance_graph} 
is the heterogeneity of the node degrees and link weights.  The numbers 
of polysemies involving individual meanings are uneven, possibly 
toward a heavy-tailed distribution (Fig.~\ref{fig:word_degree_rank}). 
This indicates concepts not only form clusters within which they are 
densely connected, but also exhibit different levels of being polysemous. 
For example, {\ttfamily EARTH/SOIL} has more than hundreds of polysemes 
while {\ttfamily SALT} has only a few. Having shown that some aspects of 
the semantic network are universal, we next ask whether the observed 
heterogeneous degrees of polysemy, possibly a manifestation of varying 
densities of near conceptual neighbors, arise as artifacts of language 
family structure in our sample, or if they are inherent to the concepts 
themselves. Simply put, is it an intrinsic property of 
the concept, {\ttfamily EARTH/SOIL}, to be extensively polysemous, 
or is it a few languages that happened to call the same concept in so many different ways.

Suppose an underlying ``universal space'' 
relative to which each language $L$ randomly draws a subset of 
polysemies for each concept $S$.  The number of polysemies 
$n_{SL}$ should then be linearly proportional to both the tendency 
of the concept to be polysemous for being close to many other concepts, 
and the tendency of the language to distinguish word senses in basic 
vocabulary.
In our network representation, a proxy for the former is
the weighted degree $n_S$ of node $S$, and a proxy for the latter is
the total weight of links $n_L$ in language $L$. 
Then the number of polysemies is expected (see Methods): 
\begin{equation}
  n_{SL}^{\mbox{\scriptsize model}} \equiv {n_S} \times
  \frac{n_L}{N} . 
\label{eq:product}
\end{equation}
This simple model indeed captures the gross features of the data very
well (Fig.~\ref{fig:productmodel_matrix} in the Appendix).  Nevertheless, 
the Kullback-Leibler divergence between the prediction 
$n_{SL}^{\mbox{\scriptsize model}}$ and the empirical data 
$n_{SL}^{\mbox{\scriptsize data}}$ 
identifies deviations beyond the sampling errors 
in three concepts---{\ttfamily MOON}, {\ttfamily SUN} and 
{\ttfamily ASHES}---that display nonlinear increase in the number of
polysemies ($p \approx 0.01$) with the tendency of the language distinguish 
word senses as Fig.~\ref{fig:saturating_words} in the Appendix shows. 
Accommodating saturation parameters (Table~\ref{tab:fitvalue}
in the Appendix) enables
the random sampling model to reproduce the empirical data in good agreement 
keeping the two parameters independent, hence
retain the universality over language groups.

\section*{Discussion}

The similarity relations between word meanings through common polysemies 
exhibit a universal structure, manifested as intrinsic closeness between 
concepts, that transcends cultural or environmental factors.  Polysemy 
arises when two or more concepts are fundamental enough to receive distinct vocabulary terms 
in some languages, yet similar enough to share a common term in others.  
The highly variable degree of these polysemies indicates such salient 
concepts are not homogeneously distributed in the {\it conceptual} 
space, and the intrinsic parameter that describes the overall 
propensity of a word to participate in polysemies can then be 
interpreted as a measure of the local density around such concept.  
Our model suggests that given the overall semantic ambiguity observed 
in the languages, such local density determines the degree of 
polysemies. 
Universal structures in lexical semantics would greatly aid another 
subject of broad interest, namely reconstruction of human phylogeny 
using linguistic data \cite{Dunn2011, Bouckaert2012}. Much progress has 
been made in reconstructing the phylogenies of word forms from known 
cognates in various languages, thanks to the ability to measure phonetic 
similarity and our knowledge of the processes of sound change. However, 
the relationship between semantic similarity and semantic shift is still 
poorly understood. The standard view in historical linguistics is that 
any meaning can change to any other meaning \cite{Hock1986, Fox1995}, and 
that no constraint is imposed on what meanings can be compared to detect 
cognates \cite{Nichols1996}.  It is, however, generally accepted among 
historical linguists that language change is gradual, and that words in 
transition from having one meaning to being extended to another meaning 
should be polysemous. If this is true, then the weights on different links 
reflect the probabilities that words in transition over these links will be
captured in ``snapshots'' by language translation at any time.  
Such semantic shifts can be modeled as diffusion in the conceptual space, 
or along a universal polysemy network where our constructed networks 
can serve an important input to methods of inferring cognates. 

The absence of significant cladistic correlation with the patterns of
polysemy suggests a possibility to extend the constructed conceptual 
space by utilizing digitally archived dictionaries of the major 
languages of the world with some confidence that their expression of 
these features is not strongly biased by correlations due to language 
family structure.  Large-corpus samples could be used to construct 
the semantic space in as yet unexplored domains using automated means.


\section*{Methods}
\subsection*{Polysemy data} 
High-quality bilingual dictionaries between the object language and the 
semantic metalanguage for cross-linguistic comparison are used to identify 
polysemies.  The 81 object languages were selected from a 
phylogenetically and geographically stratified sample of low-level 
language families or {\it genera}, 
listed in Tab. \ref{tab:languages} in the Appendex \cite{Dryer1989}.
Translations into the object language of 
each of the 22 word senses from the Swadesh basic vocabulary list were 
first obtained (See Appendix-\ref{subsec:meanings}); all translations 
(that is, all synonyms) were retained. 
Polysemies were identified by looking up the metalanguage translations 
(back-translation) of each object-language term.  The selected Swadesh 
word senses, and the selected languages are listed in the Appendix. 

We use modern European languages as a semantic metalanguage, 
{\it i.e.,\hbox{}} bilingual dictionaries 
between such languages and the other languages in our sample. This 
could be problematic if these languages themselves display polysemies; 
for example, English {\it day} expresses both {\ttfamily DAYTIME}, and 
{\ttfamily 24HR PERIOD}.  In many cases, however, the lexicographer 
is aware of these issues, and annotates the translation of the object 
language word accordingly.  In the lexical domain chosen for our study, 
standard lexicographic practice was sufficient to overcome this problem.



\subsection*{Comparing semantic networks between language groups} 
A hierarchical spectral algorithm clusters the Swadesh word senses.
Each sense $i$ is assigned 
to a position in $\mathbb{R}^n$ based on the $i$th components of the
$n$ eigenvectors of the weighted adjacency matrix.  Each eigenvector is
weighted by the square of its eigenvalue, and clustered by a greedy
agglomerative algorithm to merge the pair of clusters having the
smallest squared Euclidean distance between their centers of mass, 
through which a binary tree or {\it dendrogram} is constructed
We construct a dendrogram 
for each subgroup of languages according to nonlinguistic 
variables such as geography, topography, climate, and the presence 
or absence of a literary tradition (Table \ref{tab:groups} in Appendix).

The structural distance between the dendrograms of each pair of 
language subgroups is measured by two standard tree metrics.
The triplet distance $\dtrip$ \cite{Dobson:75,Critchlow:96} is
the fraction of the ${n \choose 3}$ distinct triplets of senses 
that are assigned a different topology in the two trees: that is, 
those for which the trees disagree as to which pair of senses are more
closely related to each other than they are to the third.  The Robinson-Foulds
distance $\drf$ \cite{Robinson-Foulds} is the number of ``cuts'' on
which the two trees disagree, where a cut is a separation of the
leaves into two sets resulting from removing an edge of the tree.

For each pair of subgroups, we perform two types of bootstrap 
experiments.  First, we compare the distance between their dendrograms 
to the distribution of distances we would see under a hypothesis that 
the two subgroups have no shared lexical structure.  Were this null 
hypothesis true, the distribution of distances would be unchanged 
under the random permutation of the senses at the leaves of each tree 
(For simplicity, the topology of the dendrograms are kept fixed.)  
Comparing the observed distance against the resulting distribution gives a 
$p$-value, called $p_1$ in Figure~\ref{fig:bootstrap}.
These $p$-values are small enough to decisively reject the null hypothesis. 
Indeed, for most pairs of groups the Robinson-Foulds distance is smaller 
than that observed in any of the 1000 bootstrap trials ($p < 0.001$) 
marked as $0^\ast$ in the table.  This gives overwhelming evidence that 
the semantic network has universal aspects that apply across language 
subgroups: for instance, in every group we tried, {\ttfamily SEA/OCEAN}, and 
{\ttfamily SALT} are more related than either is to {\ttfamily SUN}.

In the second bootstrap experiment, the null hypothesis is that the 
nonlinguistic variables have no effect on the semantic network, and 
that the differences between language groups simply result from 
random sampling: for instance, the similarity between the Americas 
and Eurasia is what one would expect from any disjoint subgroups of 
the 81 languages of given sizes 29 and 20 respectively.  To test 
this null hypothesis, we generate random pairs of 
disjoint language subgroups with the same sizes as the groups in question, 
and measure the distribution of their distances.  The $p$-values, 
called $p_2$ in Figure~\ref{fig:bootstrap}, are not small 
enough to reject this null hypothesis.  Thus, at least given the 
current data set, there is no statistical distinction between 
random sampling and empirical data 
---further supporting our thesis that it is, at least in 
part, universal.

\subsection*{Null model}
The model treats all concepts as independent members of an unbiased sample that
the aggregate summary statistics of the empirical data reflects the underlying 
structure.  The simplest model perhaps then assumes no interaction between 
concept and languages: the number of polysemies of concept $S$ in language $L$, 
that is $n_{S L}^{\mbox{\scriptsize model}}$, is linearly proportional to both 
the tendency of the concept to be polysemous and the tendency of the language 
to distinguish word senses; and these tendencies are estimated from the 
marginal distribution of the observed data as the fraction of polysemy 
associated with the concept, $p_{S}^{\mbox{\scriptsize data}}=
n_{S}^{\mbox{\scriptsize data}}/N$, and the fraction of polysemy in the 
language, $p_{S}^{\mbox{\scriptsize data}} = n_{L}^{\mbox{\scriptsize data}}/N$, 
respectively.
The model can, therefore, be expressed as, 
$p_{S L}^{\mbox{\scriptsize model}} = 
p_{S}^{\mbox{\scriptsize data}} p_{L}^{\mbox{\scriptsize data}}$, a product
of the two. 

To test the model, we compare the Kullback-Leibler (KL) divergence of 
ensembles of the model with the observation \cite{Cover1991}. 
Ensembles are generated by the multinominal distribution according to 
the probability $p_{SL}^{\mbox{\scriptsize model}}$.
The KL divergence is an appropriate measure for testing typicality of this 
random process because it is the leading exponential approximation 
(by Stirling’s formula) to the log of the multinomial distribution 
produced by Poisson sampling (see Appendix~\ref{sec:model}).
The KL divergence of ensembles is 
$
  D \left( p_{S L}^{\mbox{\scriptsize ensemble}} \right\| 
  \left. p_{S L}^{\mbox{\scriptsize model}} \right) \equiv 
  \sum_{S, L} 
  p_{S L}^{\mbox{\scriptsize ensemble}}
  \log 
  \left( 
    { p_{S L}^{\mbox{\scriptsize ensemble}} }/
    { p_{S L}^{\mbox{\scriptsize model}} }
  \right) 
$ where $p_{S L}^{\mbox{\scriptsize ensemble}}$ is the number 
of polysemies that the model generates divided by $N$, 
and the KL divergence of the empirical observation is $
  D \left( p_{S L}^{\mbox{\scriptsize data}} \right\| 
  \left. p_{S L}^{\mbox{\scriptsize model}} \right) \equiv 
  \sum_{S, L} 
  p_{S L}^{\mbox{\scriptsize data}}
  \log 
  \left( 
    { p_{S L}^{\mbox{\scriptsize data}} }/
    { p_{S L}^{\mbox{\scriptsize model}} }
  \right) 
$. Note that $ p_{S L}^{\mbox{\scriptsize data}} $ is 
${n_{SL}}^{\mbox{\scriptsize data}}/N$ and it is a different value 
from an expected value of the model, 
${n_S}^{\mbox{\scriptsize data}} {n_L^{\mbox{\scriptsize data}}}/N^2$. 
The $p$-value is the cumulative probability of 
$ D \left( p_{S L}^{\mbox{\scriptsize ensemble}} \right\| 
\left. p_{S L}^{\mbox{\scriptsize model}} \right) $ to the right of 
$ D \left( p_{S L}^{\mbox{\scriptsize data}} \right\| 
  \left. p_{S L}^{\mbox{\scriptsize model}} \right) $.


\section*{Acknowledgments}
HY acknowledges support from CABDyN Complexity Centre, and 
the support of research grants from the National Science Foundation 
(no. SMA-1312294). WC and LS acknowledge support from the University of
New Mexico Resource Allocation Committee.  TB, JW, ES, CM, and HY acknowledge 
the Santa Fe Institute, and the Evolution of Human Languages program. 
Authors thank Ilia Peiros, George Starostin, and Petter Holme for helpful 
comments. 
   W.C. and T.B. conceived of the project
   and participated in all methodological decisions. L.S. and
   W.C. collected the data, H.Y., J.W., E.S., and T.B. did the
   modeling and statistical analysis. I.M. and W.C. provided the
   cross-linguistic knowledge. H.Y., E.S., and C.M. did the network
   analysis.  The manuscript was written mainly by H.Y., E.S., W.C.,
   C.M., and T.B., and all authors agreed on the final version.

\clearpage

\section*{Appendix}

\subsection{Criteria for selection of meanings} \label{subsec:meanings}
Our translations use only lexical concepts as opposed to grammatical 
inflections or function words.  For the purpose of universality and
stability of meanings across cultures, we chose entries from the Swadesh 
200-word list of basic vocabulary.  Among these, we have selected 
categories that are likely to have single-word representation for 
meanings, and for which the referents are material entities or natural 
settings rather than social or conceptual abstractions.  We have selected 
22 words in domains concerning natural and geographic features, so that 
the web of polysemy will produce a connected graph whose structure we can 
analyze, rather than having an excess of disconnected singletons.  We have 
omitted body parts---which by the same criteria would provide a similarly 
appropriate connected domain---because these have been considered 
previously~\cite{Brown1976, Brown1979, WitkowskiBrown1978, BrownWitkowski1981}.  
The final set of 22 words are as follows:
\begin{itemize}
\item Celestial Phenomena and Related Time Units: \\
STAR, SUN, MOON, YEAR, DAY/DAYTIME, NIGHT 
\item Landscape Features: \\
SKY, CLOUD(S), SEA/OCEAN, LAKE, RIVER, MOUNTAIN
\item Natural Substances: \\
STONE/ROCK, EARTH/SOIL, SAND, ASH(ES), SALT, SMOKE, DUST, FIRE, WATER, WIND
\end{itemize}

\subsection{Language List}
The languages included in our study are listed in Tab. \ref{tab:languages}. 
Notes: Oceania includes
  Southeast Asia; the Papuan languages do not form a single
  phylogenetic group in the view of most historical linguists; other
  families in the table vary in their degree of acceptance by
  historical linguists. The classification at the genus level, which
  is of greater importance to our analysis, is generally agreed
  upon. 

\begin{table}[htp]
{\tiny
\begin{tabular}{llll}
Region & Family & Genus & Language\\
Africa & Khoisan & Northern & Ju\textbar'hoan\\
&& Central & Khoekhoegowab\\
&& Southern & !X\'o\~o\\
& Niger-Kordofanian & NW Mande & Bambara \\
&& Southern W. Atlantic & Kisi \\
&& Defoid & Yor\`ub\'a \\
&& Igboid & Igbo \\
&& Cross River & Efik \\
&& Bantoid & Swahili \\
& Nilo-Saharan & Saharan & Kanuri\\
&& Kuliak & Ik\\
&& Nilotic & Nandi\\
&& Bango-Bagirmi-Kresh & Kaba D\'em\'e\\
& Afro-Asiatic & Berber & Tumazabt\\
&& West Chadic & Hausa\\
&& E Cushitic & Rendille\\
&& Semitic & Iraqi Arabic\\
Eurasia & Basque & Basque & Basque\\
& Indo-European & Armenian & Armenian\\
&& Indic & Hindi\\
&& Albanian & Albanian\\
&& Italic & Spanish\\
&& Slavic & Russian\\
& Uralic & Finnic & Finnish\\
& Altaic & Turkic & Turkish\\
&& Mongolian & Khalkha Mongolian\\
& Japanese & Japanese & Japanese\\
& Chukotkan & Kamchatkan & Itelmen (Kamchadal)\\
& Caucasian & NW Caucasian & Kabardian\\
&& Nax & Chechen\\
& Katvelian & Kartvelian & Georgian\\
& Dravidian & Dravidian Proper & Badaga\\
& Sino-Tibetan & Chinese & Mandarin\\
&& Karen & Karen (Bwe)\\
&& Kuki-Chin-Naga & Mikir \\
&& Burmese-Lolo & Hani\\
&& Naxi & Naxi\\
Oceania & Hmong-Mien & Hmong-Mien & Hmong Njua\\
& Austroasiatic & Munda & Sora\\
&& Palaung-Khmuic & Minor Mlabri\\
&& Aslian & Semai (Sengoi)\\
& Daic & Kam-Tai & Thai\\
& Austronesian & Oceanic & Trukese\\
& Papuan & Middle Sepik & Kwoma\\
&& E NG Highlands & Yagaria\\
&& Angan & Baruya\\
&& C and SE New Guinea & Kolari\\
&& West Bougainville & Rotokas\\
&& East Bougainville & Buin\\
& Australian & Gunwinguan & Nunggubuyu\\
&& Maran & Mara\\
&& Pama-Nyungan & E and C Arrernte\\
Americas & Eskimo-Aleut & Aleut & Aleut\\
& Na-Dene & Haida & Haida\\
&& Athapaskan & Koyukon\\
& Algic & Algonquian & Western Abenaki\\
& Salishan & Interior Salish & Thompson Salish\\
& Wakashan & Wakashan & Nootka (Nuuchahnulth)\\
& Siouan & Siouan & Lakhota\\
& Caddoan & Caddoan & Pawnee\\
& Iroqoian & Iroquoian & Onondaga\\
& Coastal Penutian & Tsimshianic & Coast Tsimshian\\
&& Klamath & Klamath\\
&& Wintuan & Wintu\\
&& Miwok & Northern Sierra Miwok\\
& Gulf & Muskogean & Creek\\
& Mayan & Mayan & Itz\'a Maya\\
& Hokan & Yanan & Yana\\
&& Yuman & Cocopa\\
& Uto-Aztecan & Numic & T\"umpisa Shoshone\\
&& Hopi & Hopi\\
& Otomanguean & Zapotecan & Quiavini Zapotec\\
& Paezan & Warao & Warao\\
&& Chim\'uan & Mochica/Chimu\\
& Quechuan & Quechua & Huallaga Quechua\\
& Araucanian & Araucanian & Mapudungun (Mapuche)\\
& Tup\'i-Guaran\'i & Tup\'i-Guaran\'i & Guaran\'i\\
& Macro-Arawakan & Har\'akmbut & Amarakaeri\\
&& Maipuran & Piro\\
& Macro-Carib & Carib & Carib\\
&& Peba-Yaguan & Yagua
\end{tabular}
}
\caption{The languages included in our study. The classification 
  at the genus level, which
  is of greater importance to our analysis, is generally agreed
  upon. } \label{tab:languages}
\end{table}

\clearpage

\subsection{Language groups}
We performed several tests to see if the structure of the polysemy 
network (or whatever we're calling it) depends in a statistically 
significant way on typological features, including the presence or 
absence of a literary tradition, geography, topography, and climate.  
The typological features tested, with the numbers of languages 
indicated for each feature shown in parentheses, are listed in 
Tab. \ref{tab:groups}

\begin{table}
\begin{tabular}{|l|l|l|}
\hline 
Variable & Subset & Size \\ 
\hline
\multirow{4}{*}{Geography} & Americas & 29 \\
& Eurasia & 20 \\
& Africa & 17 \\
& Oceania & 15 \\ 
\hline
\multirow{3}{*}{Climate} & Humid & 38 \\
& Cold & 30 \\
& Arid & 13 \\
\hline
\multirow{2}{*}{Topography} & Inland & 45 \\
& Coastal & 36 \\ 
\hline
\multirow{2}{*}{Literary tradition} & Some or long literary tradition & 28 \\
& No literary tradition & 53 \\ 
\hline
\end{tabular}
\caption{Various groups of languages based on nonlinguistic variables.  For each variable we measured the difference between the subsets' semantic networks, defined as the tree distance between the dendrograms of Swadesh words generated by spectral clustering.}
\label{tab:groups}
\end{table}

\clearpage

\subsection{Model for Degree of Polysemy} \label{sec:model}

\subsubsection{Aggregation of language samples}

We now consider more formally the reasons sample aggregates may not
simply be presumed as summary statistics, because they entail implicit
generating processes that must be tested.  By demonstrating an
explicit algorithm that assigns probabilities to samples of Swadesh
node degrees, presenting significance measures consistent with the
aggregate graph and the sampling algorithm, and showing that the
languages in our dataset are typical by these measures, we justify the
use and interpretation of the aggregate graph (Fig. 2 ).

We begin by introducing an error measure appropriate to independent
sampling from a general mean degree distribution.  We then introduce
calibrated forms for this distribution that reproduce the correct
sample means as functions of both Swadesh-entry and language-weight
properties.  

The notion of consistency with random sampling is generally 
scale-dependent.  In particular, the existence of synonymous polysemy
may cause individual languages to violate criteria of randomness, but
if the particular duplicated polysemes are not correlated across
languages, even small groups of languages may rapidly converge toward
consistency with a random sample.  Therefore, we do not present 
only a single acceptance/rejection criterion for our dataset, but rather
show the smallest groupings for which sampling is consistent with 
randomness, and then demonstrate a model that reproduces the excess but 
uncorrelated synonymous polysemy within individual languages.  

\subsubsection{Independent sampling from the aggregate graph}

Figure ~\ref{fig:connectance_graph} treats all words in all 
languages as independent members of an unbiased sample.  To test the
appropriateness of the aggregate as a summary statistic, we ask: do
random samples, with link numbers equal to those in observed
languages, and with link probabilities proportional to the weights in
the aggregate graph, yield ensembles of graphs within which the actual
languages in our data are typical?
\\

{\bf Statistical tests}

The appropriate summary statistic to test for typicality, in ensembles
produced by random sampling (of links or link-ends) is the
Kullback-Leibler (KL) divergence of the sample counts from the
probabilities with which the samples were drawn~\cite{Cover1991}.
This is because the KL divergence is the leading exponential
approximation (by Stirling's formula) to the log of the multinomial
distribution produced by Poisson sampling.

The appropriateness of a random-sampling model may be tested
independently of how the aggregate link numbers are used to generate
an underlying probability model.  In this section, we will first
evaluate a variety of underlying probability models under Poisson
sampling, and then we will return to tests for deviations from
independent Poisson samples.  We first introduce notation: For a 
single language, the relative degree (link frequency), which is used  
as the normalization of a probability, is denoted as $p_{S \mid
  L}^{\mbox{\scriptsize data}} \equiv n_S^L / n^L$, and for the joint
configuration of all words in all languages, the link frequency of a
single entry relative to the total $N$ will be denoted $p_{S
  L}^{\mbox{\scriptsize data}} \equiv n_S^L / N = \left( n_S^L / n^L
\right) \left( n^L / N \right) \equiv p_{S \mid L}^{\mbox{\scriptsize
    data}} p_L^{\mbox{\scriptsize data}}$.

Corresponding to any of these, we may generate samples of links to
define the null model for a random process, which we denote
${\hat{n}}_S^L$, ${\hat{n}}^L$, etc.  We will generally use samples
with exactly the same number of total links $N$ as the data.  The
corresponding sample frequencies will be denoted by $p_{S \mid
  L}^{\mbox{\scriptsize sample}} \equiv {\hat{n}}_S^L / {\hat{n}}^L$
and $p_{S L}^{\mbox{\scriptsize sample}} \equiv {\hat{n}}_S^L / N =
\left( {\hat{n}}_S^L / {\hat{n}}^L \right) \left( {\hat{n}}^L / N
\right) \equiv p_{S \mid L}^{\mbox{\scriptsize sample}}
p_L^{\mbox{\scriptsize sample}}$, respectively.

Finally, the calibrated model, which we define from properties of
aggregated graphs, will be the prior probability from which samples
are drawn to produce $p$-values for the data.  We denote the model
probabilities (which are used in sampling as ``true'' probabilities
rather than sample frequencies) by $p_{S \mid
  L}^{\mbox{\scriptsize model}}$, $p_{SL}^{\mbox{\scriptsize model}}$,
and $p_L^{\mbox{\scriptsize model}}$.

For $n^L$ links sampled independently from the distribution $p_{S \mid
  L}^{\mbox{\scriptsize sample}}$ for language $L$, the multinomial
probability of a particular set $\left\{ n_S^L \right\}$ may be
written, using Stirling's formula to leading exponential order, as
\begin{equation}
  p \! 
  \left( 
    \left\{ n_S^L \right\} \mid n^L
  \right) \sim 
  e^{- n^L
    D \left( p_{S \mid L}^{\mbox{\scriptsize sample}} \right\| 
    \left. p_{S \mid L}^{\mbox{\scriptsize model}} \right)
  }
\label{eq:prob_sample_L}
\end{equation}
where the Kullback-Leibler (KL) divergence~\cite{Cover1991}
\begin{equation}
  D \left( p_{S \mid L}^{\mbox{\scriptsize sample}} \right\| 
  \left. p_{S \mid L}^{\mbox{\scriptsize model}} \right) \equiv 
  \sum_S 
  p_{S \mid L}^{\mbox{\scriptsize sample}}
  \log 
  \left( 
    \frac{
      p_{S \mid L}^{\mbox{\scriptsize sample}}
    }{
      p_{S \mid L}^{\mbox{\scriptsize model}}
    }
  \right) . 
\label{eq:D_KL_L}
\end{equation}
For later reference, note that the leading quadratic approximation to
Eq.~(\ref{eq:D_KL_L}) is 
\begin{equation}
  n^L 
  D \left( p_{S \mid L}^{\mbox{\scriptsize sample}} \right\| 
  \left. p_{S \mid L}^{\mbox{\scriptsize model}} \right) \approx 
  \frac{1}{2}
  \sum_S 
  \frac{
    {
      \left( 
        {\hat{n}}_S^L  - n^L p_{S \mid L}^{\mbox{\scriptsize model}} 
      \right) 
    }^2 
  }{
    n^L p_{S \mid L}^{\mbox{\scriptsize model}} 
  } , 
\label{eq:D_KL_var_L}
\end{equation}
so that the variance of fluctuations in each word is proportional to
its expected frequency. 

As a null model for the joint configuration over all languages in our
set, if $N$ links are drawn independently from the distribution $p_{S
  L}^{\mbox{\scriptsize sample}}$, the multinomial probability of a
particular set $\left\{ n_S^L \right\}$ is given by
\begin{equation}
  p \! 
  \left( 
    \left\{ n_S^L \right\} \mid N \right) \sim 
  e^{- N
    D \left( p_{S L}^{\mbox{\scriptsize sample}} \right\| 
    \left. p_{S L}^{\mbox{\scriptsize model}} \right)
  }
\label{eq:prob_sample_joint}
\end{equation}
where\footnote{As long as we calibrate $p_L^{\mbox{\scriptsize
      model}}$ to agree with the per-language link frequencies $n^L /
  N$ in the data, the data will always be counted as more typical than
  almost-all random samples, and its probability will come entirely
  from the KL divergences in the individual languages.}
\begin{eqnarray}
\lefteqn{
  D \left( p_{S L}^{\mbox{\scriptsize sample}} \right\| 
  \left. p_{S L}^{\mbox{\scriptsize model}} \right) \equiv 
  \sum_{S, L} 
  p_{S L}^{\mbox{\scriptsize sample}}
  \log 
  \left( 
    \frac{
      p_{S L}^{\mbox{\scriptsize sample}}
    }{
      p_{S L}^{\mbox{\scriptsize model}}
    }
  \right) 
} & & 
\nonumber \\
& = & 
  D \left( p_L^{\mbox{\scriptsize sample}} \right\| 
  \left. p_L^{\mbox{\scriptsize model}} \right) + 
  \sum_L 
  p_L^{\mbox{\scriptsize sample}}
  D \left( p_{S \mid L}^{\mbox{\scriptsize sample}} \right\| 
  \left. p_{S \mid L}^{\mbox{\scriptsize model}} \right)\! . 
\nonumber \\
\label{eq:D_KL_joint}
\end{eqnarray}

Multinomial samples of assignments ${\hat{n}}_S^L$ to each of the $22
\times 81$ $\left( \mbox{Swadesh}, \mbox{Language} \right)$ pairs,
with $N$ links total drawn from distribution
${p_S^L}^{\mbox{\scriptsize null}}$, will produce KL divergences
uniformly distributed in the coordinate $d\Phi \equiv e^{-D_{KL}}
dD_{KL}$, corresponding to the uniform increment of cumulative
probability in the model distribution.  We may therefore use the
cumulative probability to the right of $D \left( p_{S
    L}^{\mbox{\scriptsize data}} \right\| \left. p_{S
    L}^{\mbox{\scriptsize model}} \right)$ (one-sided $p$-value), in
the distribution of samples ${\hat{n}}_S^L$, as a test of consistency
of our data with the model of random sampling.

In the next two subsections we will generate and test candidates for
$p^{\mbox{\scriptsize model}}$ which are different functions of the
mean link numbers on Swadesh concepts and the total links numbers in
languages.
\\

{\bf Product model with intrinsic property of concepts}

In general we wish to consider the consistency of joint configurations
with random sampling, as a function of an aggregation scale.  To do
this, we will rank-order languages by increasing $n^L$, form
non-overlapping bins of 1, 3, or 9 languages, and test the resulting
binned degree distributions against different mean-degree and sampling
models.  We denote by $\left< n^L \right>$ the average total link
number in a bin, and by $\left< n_S^L \right>$ the average link number
per Swadesh entry in the bin.  The simplest model, which assumes no
interaction between concept and language properties, makes the model
probability $p_{SL}^{\mbox{\scriptsize model}}$ a product of its
marginals.  It is estimated from data without regard to binning, as
\begin{equation}
  p_{SL}^{\mbox{\scriptsize product}} 
\equiv 
  \frac{n_S}{N} \times 
  \frac{n^L}{N} . 
\label{Seq:product}
\end{equation}
The $22 \times 81$ independent mean values are thereby specified in
terms of $22 + 81$ sample estimators.

\begin{figure}[t]
  \begin{center} 
  \includegraphics[scale=0.5]{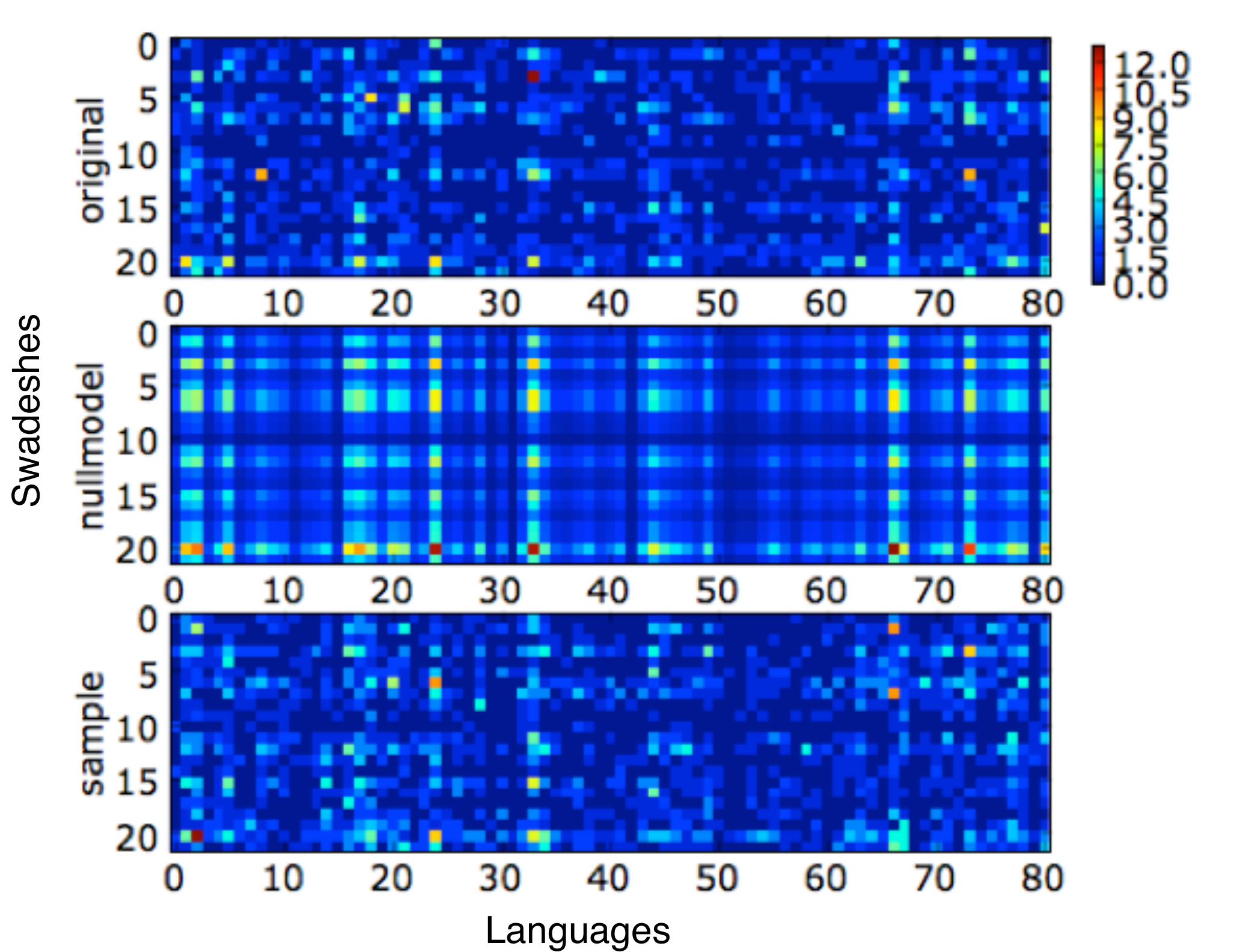}
  \caption{  
  Plots for the data $n_S^L$, $Np^{product}_{SL}$, $n_{SL}^{sample}$
  in accordance with Fig. 4S (f).
  The colors denote corresponding numbers of the scale. 
  The original data in the first panel with the sample in the last panel
  seems to agree reasonably well.
    \label{fig:productmodel_matrix} 
  }
  \end{center}
\end{figure}

The KL divergence of the joint configuration of links in the actual
data from this model, under whichever binning is used, becomes 
\begin{equation}
  D \left( p_{S L}^{\mbox{\scriptsize data}} \right\| 
  \left. p_{S L}^{\mbox{\scriptsize model}} \right) = 
  D\left( \frac{\left< n_S^L \right>}{N} \left\| 
  \frac{n_S}{N}
  \frac{\left< n^L \right>}{N} \right. \right)
\label{eq:Dkb_product}
\end{equation}
As we show in Fig.~\ref{fig:sample_KL_hist_data_9_joint} below, even for
9-language bins which we expect to average over a large amount of
language-specific fluctuation, the product model is ruled out at the
$1\%$ level.  

We now show that a richer model, describing interaction between word
and language properties, accepts not only the 9-language aggregate,
but also the 3-language aggregate with a small adjustment of the 
language size to which words respond (to produce consistency with 
word-size and language-size marginals).  Only fluctuation statistics 
at the level of the joint configuration of 81 individual languages 
remains strongly excluded by the null model of random sampling.
\\

{\bf Product model with saturation} 

An inspection of the deviations of our data from the product model shows 
that the initial propensity of a word to participate in polysemies, as 
inferred in languages where that word has few links, in general 
overestimates the number of links (degree).  Put it differently, 
languages seem to place limits on the weight of single polysemies, 
favoring distribution over distinct polysemies, but the number of 
potential distinct polysemies is an independent parameter from the 
likelihood that the available polysemies will be formed.  Interpreted 
in terms of our supposed semantic space, the proximity of target words 
to a Swadesh entry may determine the likelihood that they will be
polysemous with it, but the total number of proximal targets may vary
independently of their absolute proximity.  These limits on the 
number of neighbors of each concept are captured by additional 22 
parameters. 

\begin{figure}[t]
  \begin{center} 
  \includegraphics[scale=0.7]{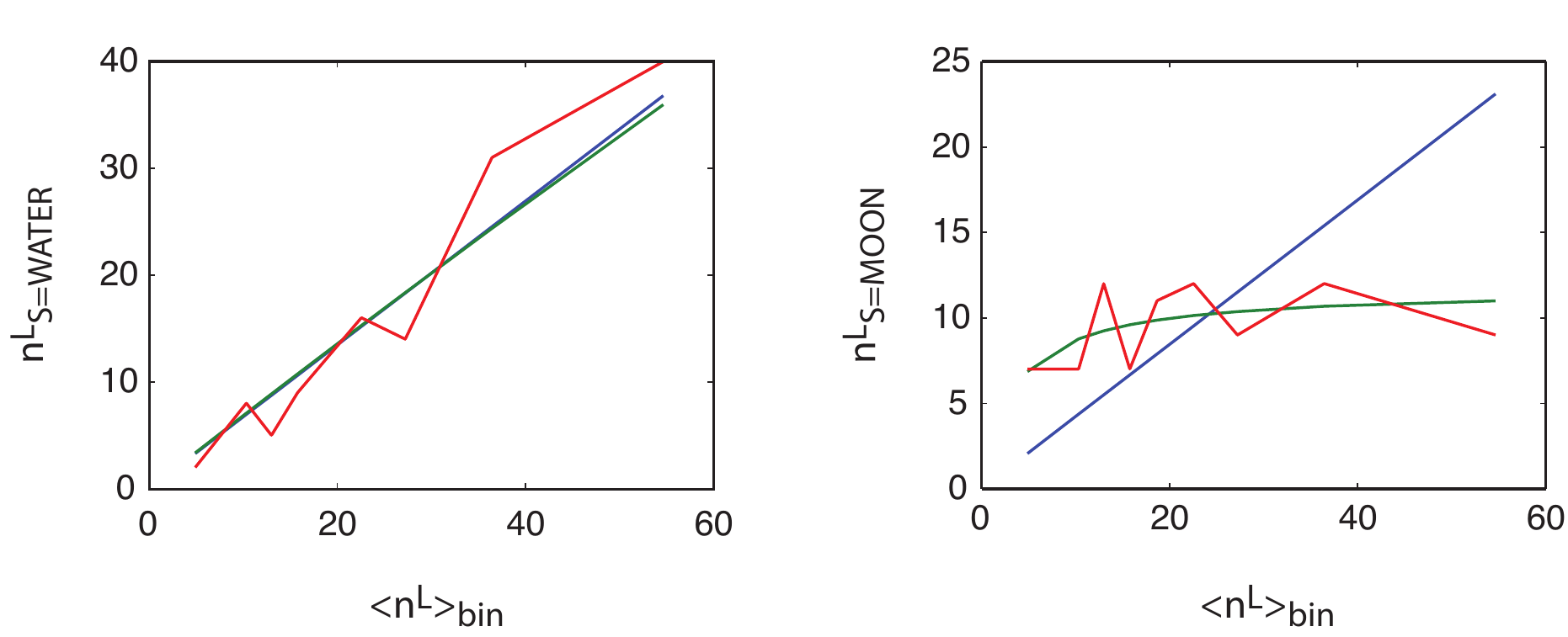}
  \caption{  
    Plots of the saturating function~(\ref{eq:sat2}) with the
    parameters given in Table~\ref{tab:fitvalue}, compared to $\left<
      n_S^L \right>$ (ordinate) in 9-language bins (to increase sample
    size), versus bin-averages $\left< n^L \right>$ (abscissa).  Red
    line is drawn 
    through data values, blue is the product model, and green is the
    saturation model.  WATER requires
    no significant deviation from the product model ($B_{\rm WATER} / N
    \gg 20$), while MOON shows the lowest saturation value
    among the Swadesh entries, at $B_{\rm MOON} \approx 3.4$.  
    \label{fig:saturating_words} 
  }
  \end{center}
\end{figure}

\begin{table}[ht]
  \begin{tabular}{| l | r | r |}
  \hline
  Meaning category & Saturation: $ B_S$ & Propensity ${\tilde{p}}_S$ \\ 
  \hline
        STAR & $ 1234.2$ & 0.025 \\ \hline
        SUN  & $25.0$ & 0.126 \\ \hline
        YEAR  & $1234.2$ & 0.021 \\ \hline
        SKY  & $1234.2$ & 0.080 \\ \hline
        SEA/OCEAN  & $1234.2$ & 0.026 \\ \hline
        STONE/ROCK  & $1234.2$ & 0.041 \\ \hline
        MOUNTAIN  & $1085.9$ & 0.049 \\ \hline
        DAY/DAYTIME  & $195.7$ & 0.087 \\ \hline
        SAND  & $1234.2$ & 0.026 \\ \hline
        ASH(ES)  & $13.8$ & 0.068 \\ \hline
        SALT  & $1234.2$ & 0.007 \\ \hline
        FIRE  & $1234.2$ & 0.065 \\ \hline
        SMOKE  & $1234.2$ & 0.031 \\ \hline
        NIGHT  & $89.3$ & 0.034 \\ \hline
        DUST  & $246.8$ & 0.065 \\ \hline
        RIVER  & $336.8$ & 0.048 \\ \hline
        WATER  & $1234.2$ & 0.073 \\ \hline
        LAKE  & $1234.2$ & 0.047 \\ \hline
        MOON  & $1.2$ & 0.997 \\ \hline
        EARTH/SOIL  & $1234.2$ & 0.116 \\ \hline
        CLOUD(S)  & $53.4$ & 0.033 \\ \hline
        WIND  & $1234.2$ & 0.051 \\
  \hline
  \end{tabular}
  \caption{
    A table of fitted values of parameters $B_S$ and ${\tilde{p}}_S$
    for the saturation model of Eq.~(\ref{eq:sat2}) . The saturation
    value $B_S$ is an asymtotic number of meanings associated with the
    entry $S$, and the propensity ${\tilde{p}}_S$ is a rate at which
    the number of polysemies increases with $n^L$ at low $n_S^L$.
    \label{tab:fitvalue}
  } 
\end{table}

To accommodate such characteristic, we revise the model Eq. 
(\ref{Seq:product}) to the following function:
\begin{displaymath}
  \frac{A_S \left< n^L \right>}{B_S + \left< n^L \right>} . 
\end{displaymath}
where degree numbers $\left< n_S^L \right>$ for each Swadesh $S$ is 
proportional to $A_S$ and language size, but is bounded by $B_S$, 
the number of proximal concepts.  The corresponding model probability 
for each language then becomes 
\begin{eqnarray}
  p_{SL}^{\mbox{\scriptsize sat}} = 
  \frac{(A_S/B_S)( n^L/N)}{1 + n^L/B_S} 
& \equiv &  
  \frac{
    {\tilde{p}}_S p_L^{\mbox{\scriptsize data}}
  }{
    1+p_L^{\mbox{\scriptsize data}} N/ B_S
  } 
. 
\label{eq:sat2}
\end{eqnarray}
As all $B_S / N \rightarrow \infty$ we recover the product model, with
$p_L^{\mbox{\scriptsize data}} \equiv n^L / N$ and ${\tilde{p}}_S
\rightarrow n_S / N$.  

A first-level approximation to fit parameters $A_S$ and $B_S$ is given
by minimizing the weighted mean-square error
\begin{equation}
  E \equiv 
  \sum_L
  \frac{1}{\left< n^L \right>}
  \sum_S
  {
    \left( 
      \left< n_S^L \right> - 
      \frac{
        A_S \left< n^L \right>
      }{
        B_S + \left< n^L \right>
      }
    \right) 
  }^2 . 
\label{eq:error_sat}
\end{equation}
The function~(\ref{eq:error_sat}) assigns equal penalty to squared
error within each language bin $\sim \left< n^L \right>$, proportional
to the variance expected from Poisson sampling.  The fit values
obtained for $A_S$ and $B_S$ do not depend sensitively on the size of
bins except for the Swadesh entry MOON in the case where all 81
single-language bins are used.  MOON has so few polysemies, but the
MOON/month polysemy is so likely to be found, that the language
Itelman, with only one link, has this polysemy.  This point leads to
instabilities in fitting $B_{\rm MOON}$ in single-language bins.  For
bins of size 3--9 the instability is removed.  Representative fit
parameters across this range are shown in Table~\ref{tab:fitvalue}.  
Examples of the saturation model for two words, plotted
against the 9-language binned degree data in
Fig.~\ref{fig:saturating_words}, show the range of behaviors spanned
by Swadesh entries.

\begin{figure}[t]
  \begin{center} 
  \includegraphics[scale=0.45]{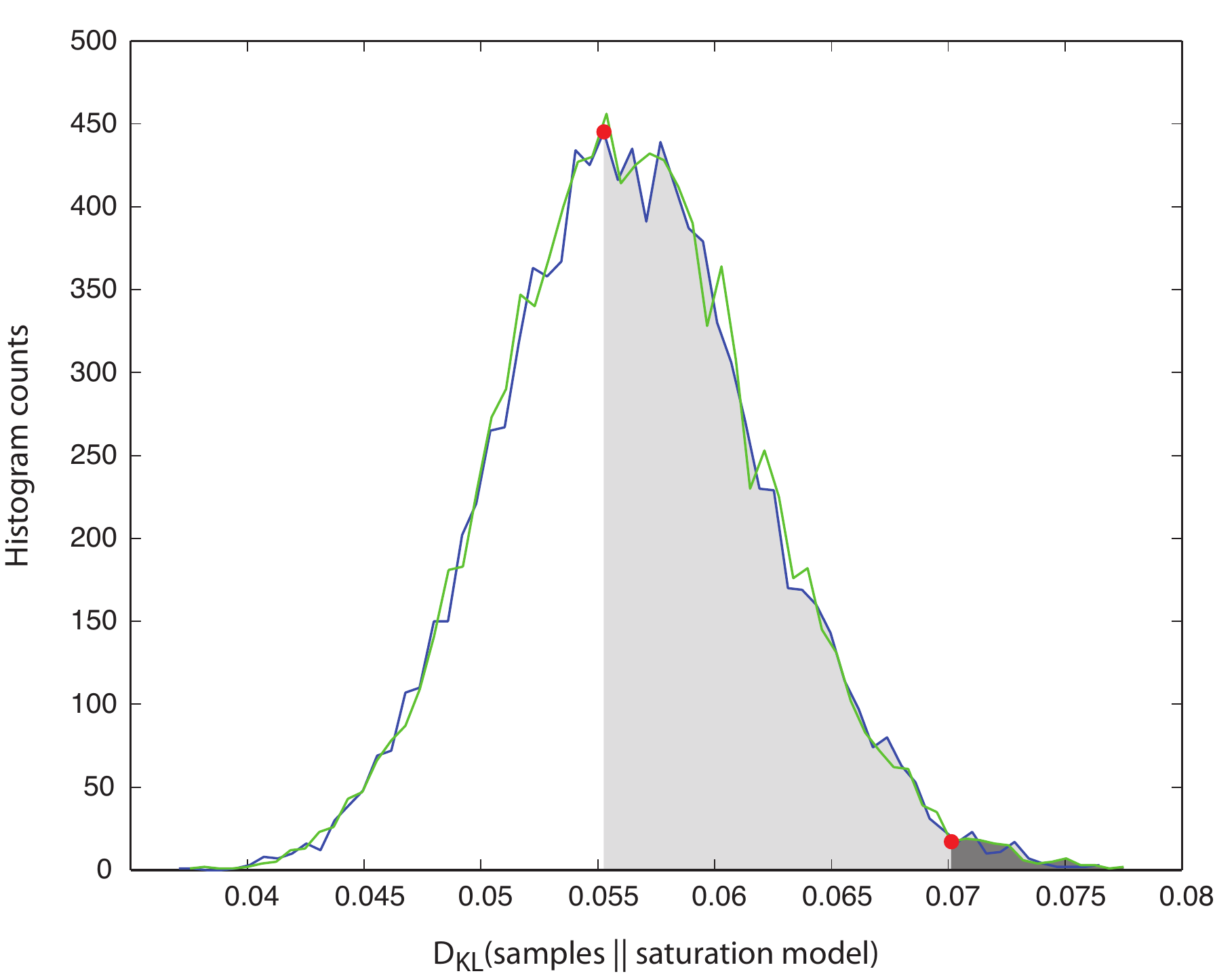}
  \caption{
    Kullback-Leibler divergence of link frequencies in our data,
    grouped into non-overlapping 9-language bins ordered by rank, from
    the product distribution~(\ref{Seq:product}) and the saturation
    model~(\ref{eq:sat2}).  Parameters $A_S$ and $B_S$ have been
    adjusted (as explained in the text) to match 
    the word- and language-marginals.  From 10,000 random samples
    ${\hat{n}}_S^L$, (green) histogram for the
    product model; (blue) histogram for the saturation model;  
    (red dots) data.  The product model rejects the 9-language
    joint binned configuration at the at $1\%$ level (dark shading),
    while the saturation model is typical of the same configuration 
		at $\sim 59\%$ (light shading).  
    \label{fig:sample_KL_hist_data_9_joint} 
  }
  \end{center}
\end{figure}

\begin{figure}[ht]
  \begin{center} 
  \includegraphics[scale=0.4]{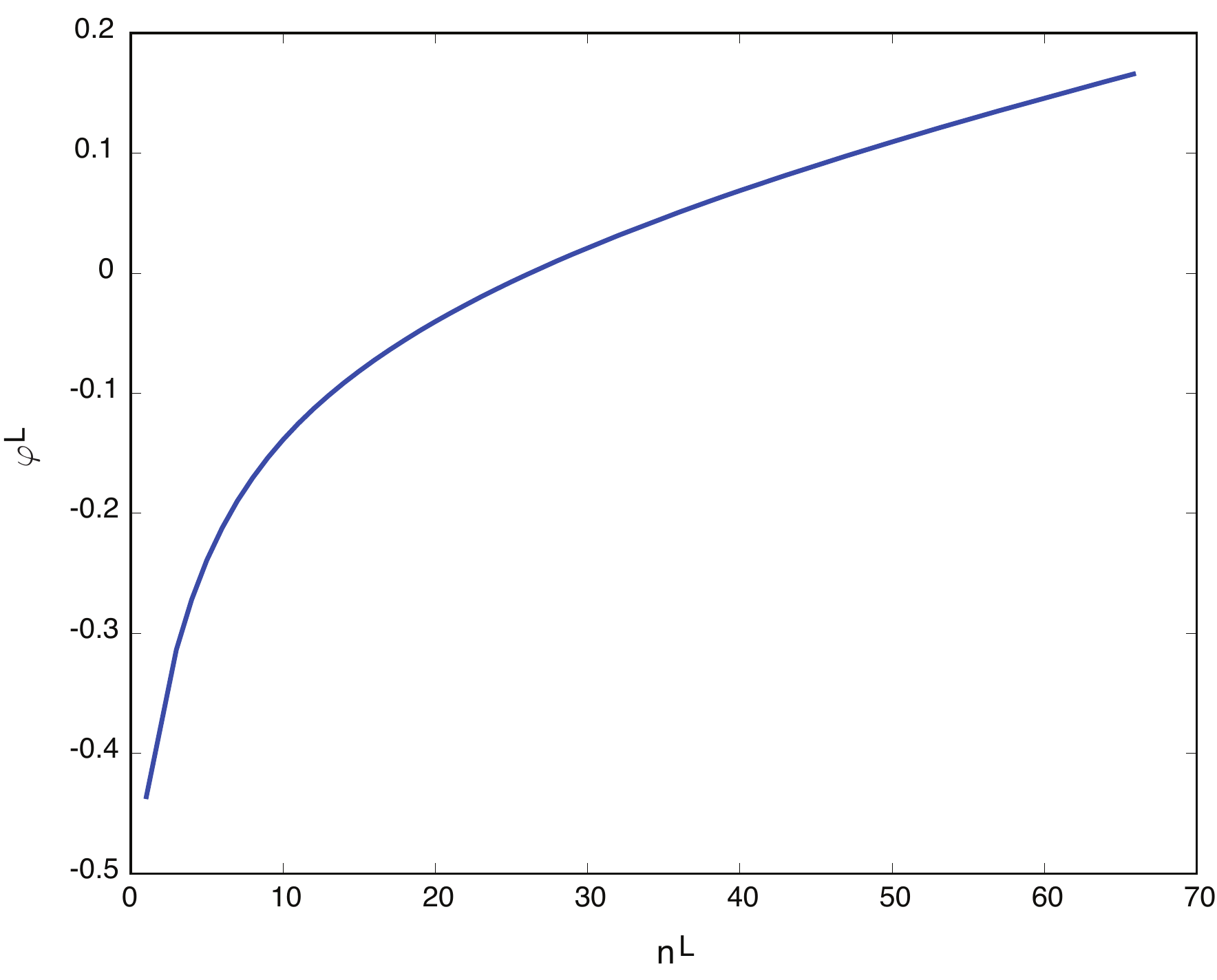}
  \caption{
    Plot of the correction factor ${\varphi}^L$ versus $n^L$ for
    individual languages in the probability model used in text, with
    parameters $B_S$ and ${\tilde{p}}_S$ shown in
    Table~\ref{tab:fitvalue}.  Although ${\varphi}^L$ values were
    individually solved with Newton's method to ensure that the
    probability model matched the whole-language link values, the
    resulting correction factors are a smooth function of $n^L$.  
    \label{fig:varphis_vs_nLs} 
  }
  \end{center}
\end{figure}

The least-squares fits to $A_S$ and $B_S$ do not directly yield a
probability model consisent with the marginals for language size that,
in our data, are fixed parameters rather than sample variables to be
explained.  They closely approximate the marginal N $\sum_L
p_{SL}^{\mbox{\scriptsize sat}} \approx n_S$ (deviations $< 1$ link
for every $S$) but lead to mild violations $N \sum_S
p_{SL}^{\mbox{\scriptsize sat}} \neq n^L$.  We corrected for this by
altering the saturation model to suppose that, rather than word
properties' interacting with the exact value $n^L$, they interact with
a (word-independent but language-dependent) multiplier $\left( 1 +
  {\varphi}^L \right) n_L$, so that the model for $n_S^L$ in each
language becomes becomes
\begin{displaymath}
  \frac{
    A_S \left( 1 + {\varphi}^L \right) n^L 
  }{
    B_S + \left( 1 + {\varphi}^L \right) n^L 
  } , 
\end{displaymath}
in terms of the least-squares coefficients $A_S$ and $B_S$ of
Table~\ref{tab:fitvalue}.  The values of ${\varphi}^L$ are solved with
Newton's method to produce $N \sum_S p_{SL}^{\mbox{\scriptsize sat}}
\rightarrow n^L$, and we checked that they preserve $N \sum_L
p_{SL}^{\mbox{\scriptsize sat}} \approx n_S$ within small fractions of
a link.  The resulting adjustment parameters are plotted versus $n^L$
for individual languages in Fig.~\ref{fig:varphis_vs_nLs}.  
Although they were computed individually for each $L$, they
form a smooth function of $n^L$, possibly suggesting a refinement of
the product model, but also perhaps reflecting systematic interaction
of small-language degree distributions with the error
function~(\ref{eq:error_sat}).  

\begin{figure}[t]
  \begin{center} 
  \includegraphics[scale=0.45]{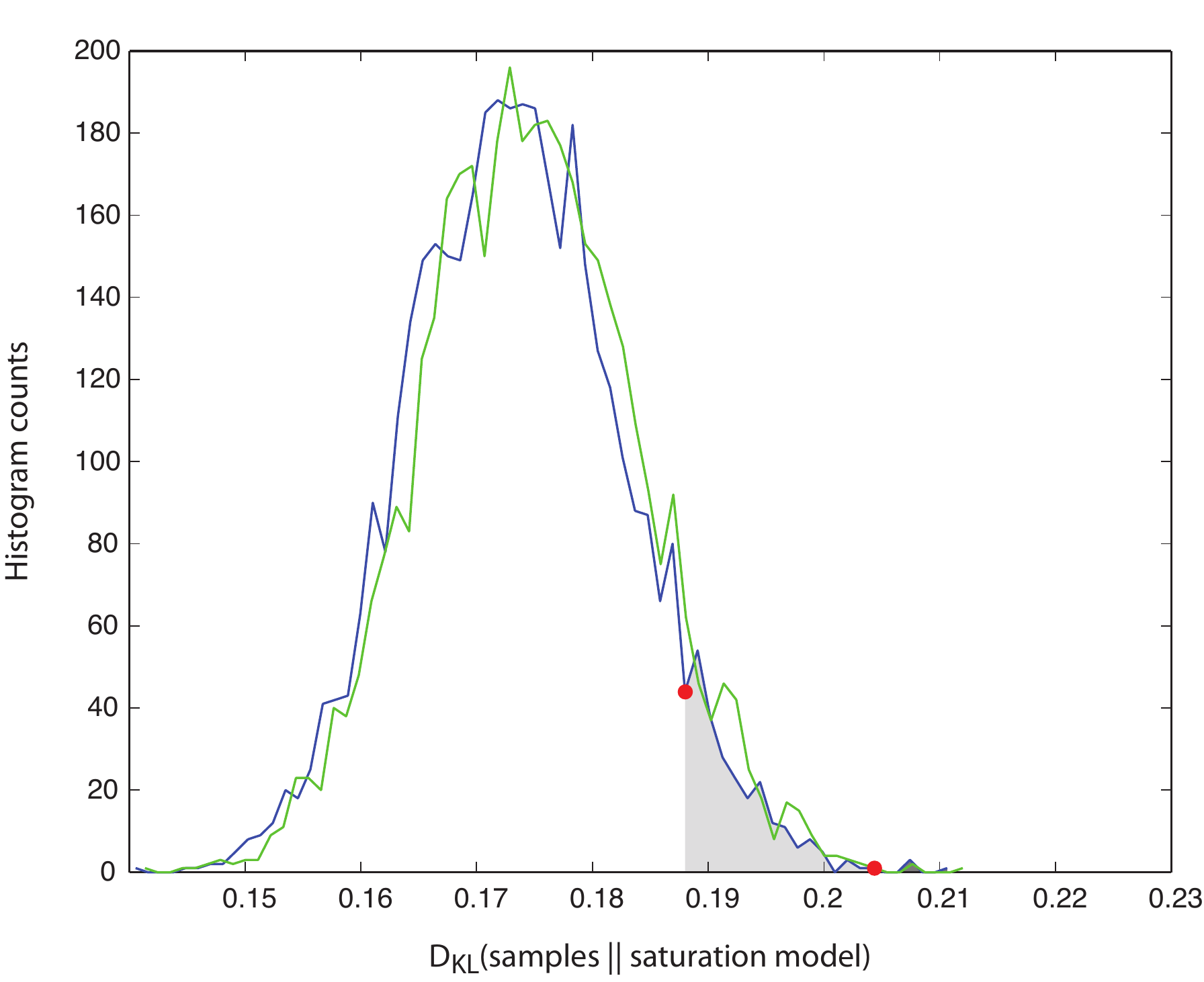}
  \caption{
    The same model parameters as in
    Fig.~\ref{fig:sample_KL_hist_data_9_joint} is now marginally 
    plausible for the joint configuration of 27 three-language bins 
    in the data, at the $7\%$ level (light shading).  For reference, 
    this fine-grained joint configuration rejects the null model of 
    independent sampling from the product model at 
    $p-value \approx {10}^{-3}$ (dark shading in the extreme tail).  
    4000 samples were used to generate this test distribution. The 
    blue histogram is for the saturation model, the green histogram 
    for the product model, and the red dots are generated data.
    \label{fig:sample_KL_hist_data_27_joint} 
  }
  \end{center}
\end{figure}

With the resulting joint distribution $p_{SL}^{\mbox{\scriptsize
    sat}}$, tests of the joint degree counts in our dataset for
consistency with multinomial sampling in 9 nine-language bins are
shown in Fig.~\ref{fig:sample_KL_hist_data_9_joint}, and results of
tests using 27
three-language bins are shown in
Fig.~\ref{fig:sample_KL_hist_data_27_joint}.  Binning nine languages
clearly averages over enough language-specific variation to make the
data strongly typical of a random sample ($P \sim 59\%$), while the
product model (which also preserves marginals) is excluded at the
$1\%$ level.  The marginal acceptance of the data even for the joint
configuration of three-language bins ($P \sim 7\%$) suggests that
language size $n^L$ is an excellent explanatory variable and that
residual language variations cancel to good approximation even in
small aggregations.

\subsubsection{Single instances as to aggregate representation} 

The preceding subsection showed intermediate scales of
aggregation of our language data are sufficiently random that they can
be used to refine probability models for mean degree as a function of
parameters in the globally-aggregated graph.  The saturation model,
with data-consistent marginals and multinomial sampling, is weakly
plausible by bins of as few as three languages.  Down to this scale, we
have therefore not been able to show a requirement for deviations from
the independent sampling of links entailed by the use of the aggregate
graph as a summary statistic.  However, we were unable to find a
further refinement of the mean distribution that would reproduce the
properties of single language samples.  In this section we
characterize the nature of their deviation from independent samples of
the saturation model, show that it may be reproduced by models of
non-independent (clumpy) link sampling, and suggest that these reflect
excess synonymous polysemy.\\

{\bf Power tests and uneven distribution of single-language $p$-values}

\begin{figure}[ht]
  \begin{center} 
  \includegraphics[scale=0.45]{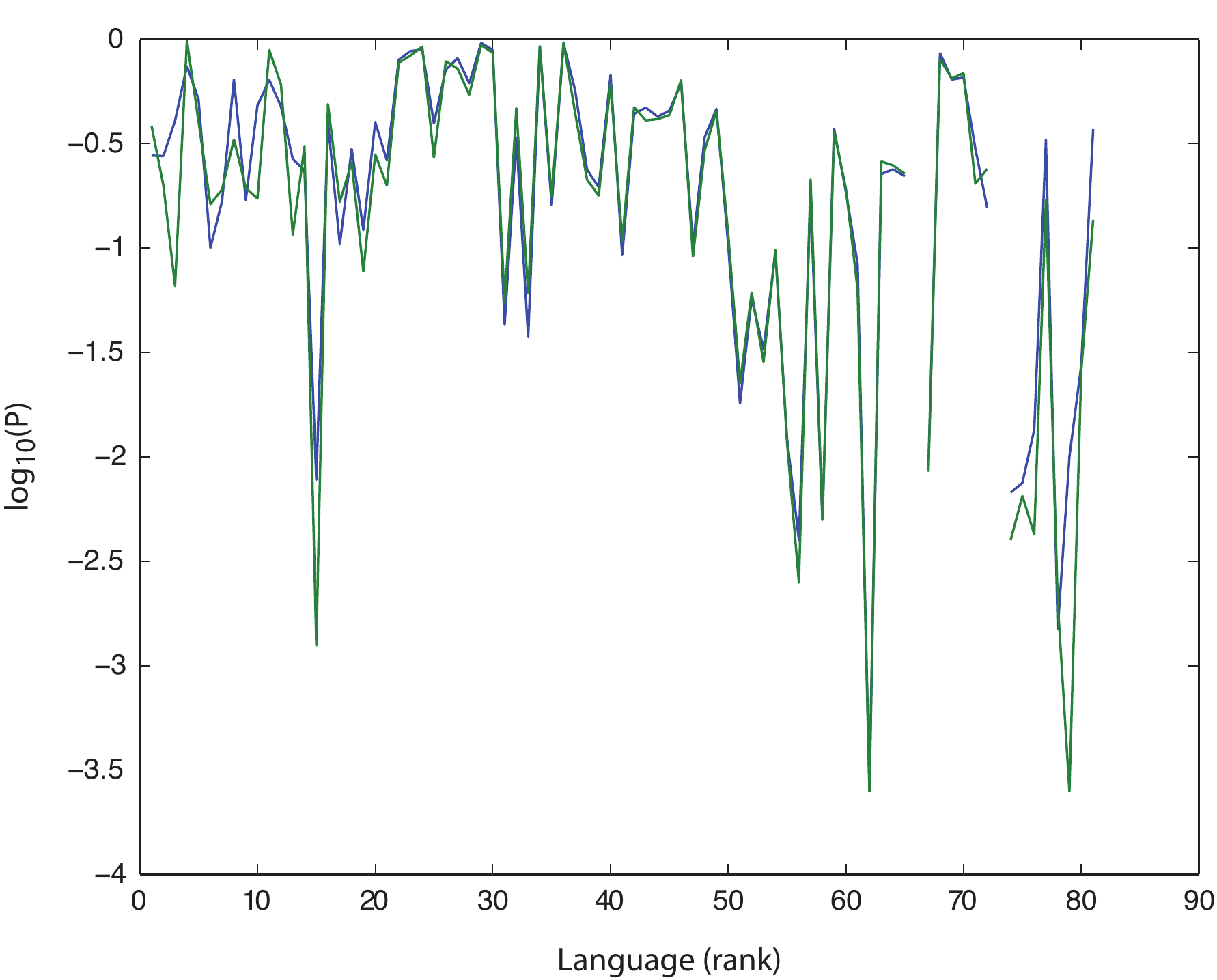}
  \caption{
  %
  %
  $\log_{10} \! \left( p\rm{-value} \right)$ by KL divergence, relative to
  4000 random samples per language, plotted versus language rank in
  order of increasing $n^L$.  Product model (green) shows equal or
  lower $p$-values for almost all languages than the saturation model
  (blue).  Three languages -- Basque, Haida, and Yor\`ub\'a -- had value
  $p = 0$ consistently across samples in both models, and are removed
  from subsequent regression estimates.  A trend toward decreasing $p$
  is seen with increase in $n^L$.
    \label{fig:one_lang_p_vals} 
  }
  \end{center}
\end{figure}

\begin{figure}[ht]
  \begin{center} 
  \includegraphics[scale=0.45]{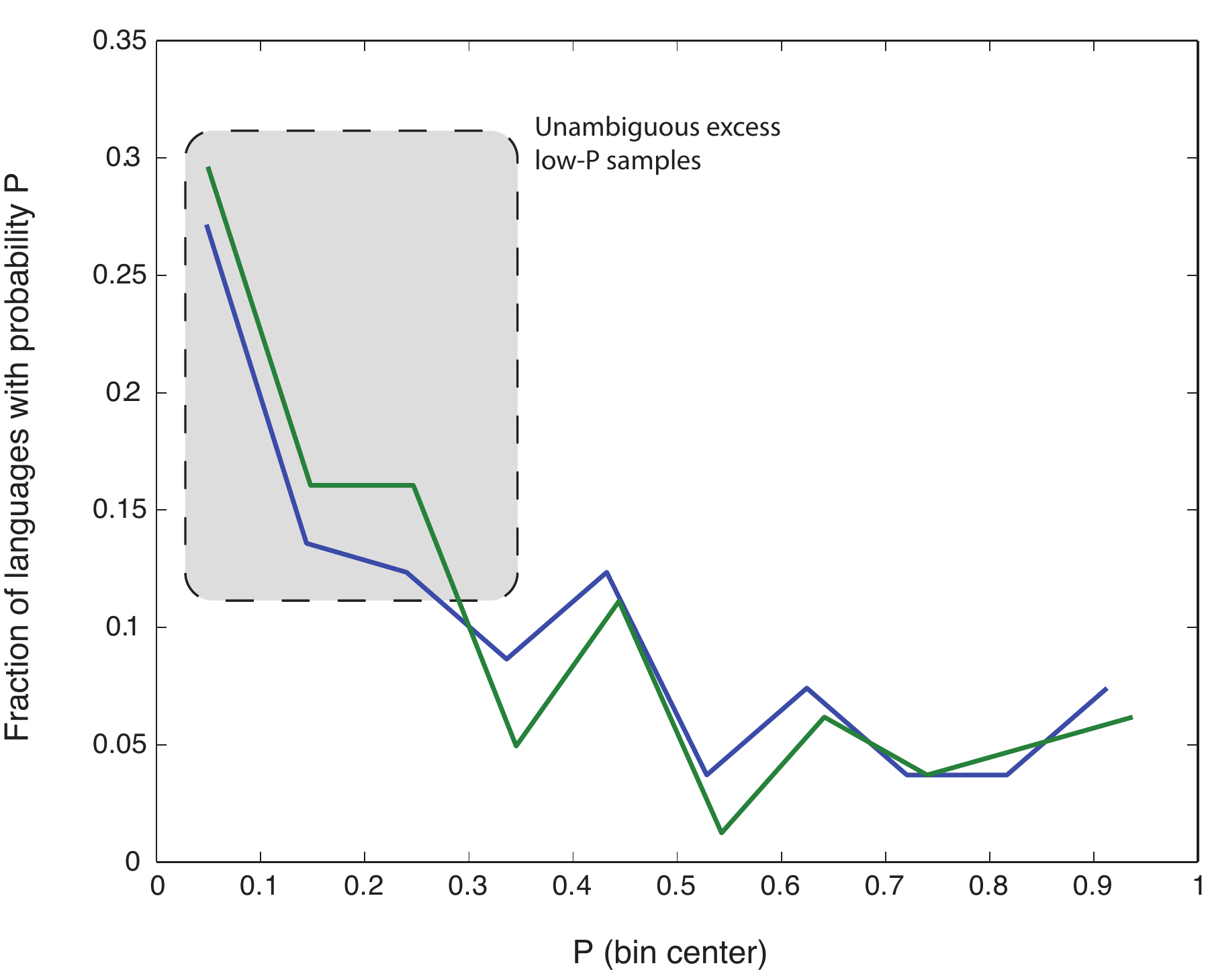}
  \includegraphics[scale=0.45]{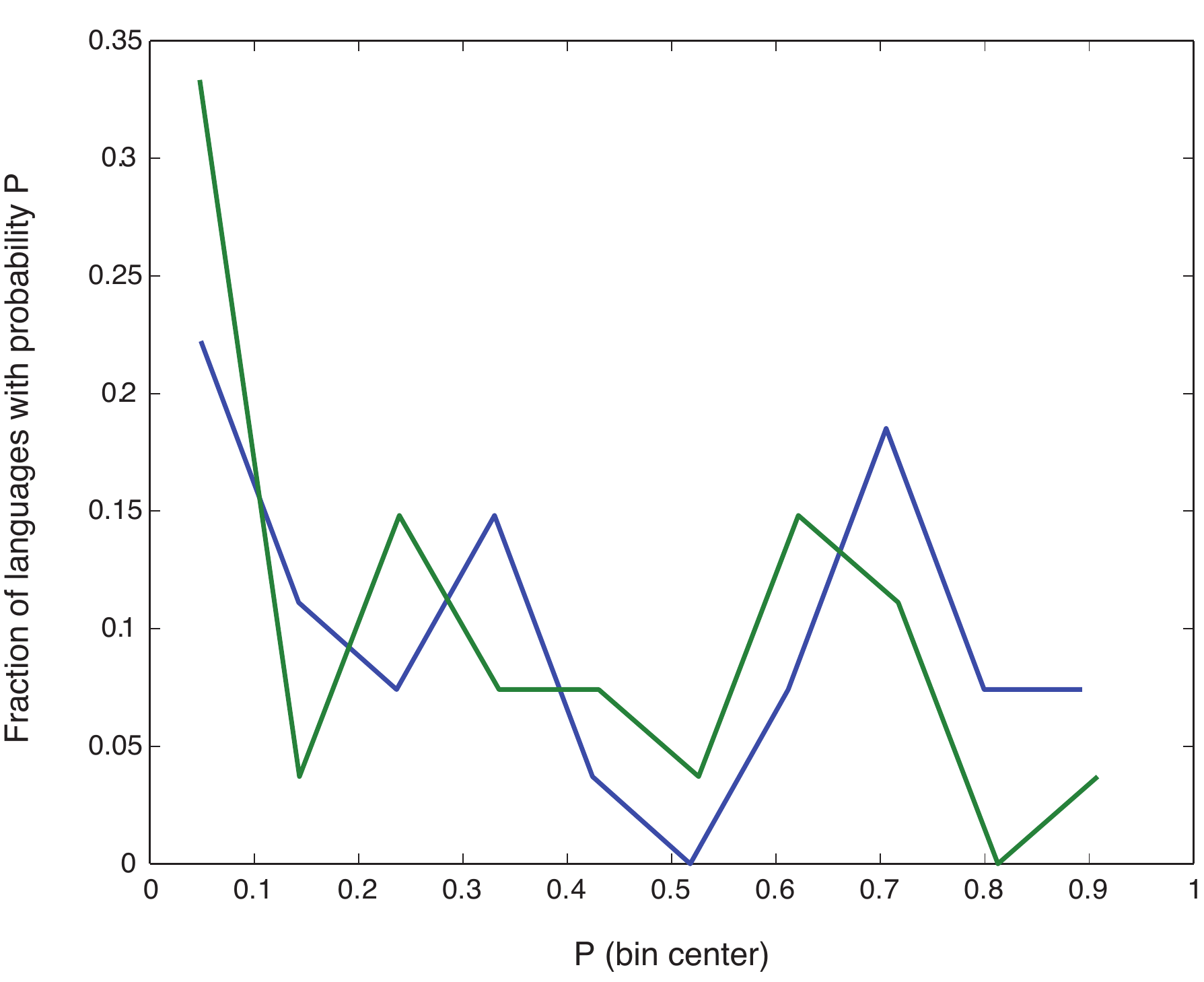} \\
  \caption{
  (Upper panel) Normalized histogram of $p$-values from the 81
  languages plotted in Fig.~\ref{fig:one_lang_p_vals}.  The saturation
  model (blue) produces a fraction $\sim 0.05 \times 81 \approx 4-5$
  languages in the lowest $p$-values $\left\{ 0.05, 0.1 \right\}$
  above the roughly-uniform background for the rest of the interval
  (shaded area with dashed boundary).
  A further excess of 2--3 languages with $p$-values in the range
  $\left[ 0 , 0.2 \right]$ for the product model (green) reflects the
  part of the mismatch corrected through mean values in the saturation
  model.  (Lower panel) Corresponding histogram of $p$-values for 27
  three-language aggregate degree distributions.  Saturation model
  (blue) is now marginally consistent with a uniform distribution,
  while the product model (green) still shows slight excess of low-$p$
  bins.  Coarse histogram bins have been used in both panels to
  compensate for small sample numbers in the lower panel, while
  producing comparable histograms.
    \label{fig:p_dists_marg_sat} 
  }
  \end{center}
\end{figure}

To evaluate the contribution of individual languages versus language
aggregates to the acceptance or rejection of random-sampling models,
we computed $p$-values for individual languages or language bins,
using the KL-divergence~(\ref{eq:D_KL_L}).  A plot of the
single-language $p$-values for both the null (product) model and the
saturation model is shown in Fig.~\ref{fig:one_lang_p_vals}.  
Histograms for both single languages (from the values in
Fig.~\ref{fig:one_lang_p_vals}) and aggregate samples formed by
binning consecutive groups of three languages are shown in
Fig.~\ref{fig:p_dists_marg_sat}.

For samples from a random model, $p$-values would be uniformly
distributed in the unit interval, and histogram counts would have a
multinomial distribution with single-bin fluctuations depending on the
total sample size and bin width.  Therefore, Fig.~\ref{fig:p_dists_marg_sat}
provides a power test of our summary statistics.  The variance of the
multinomial may be estimated from the large-$p$-value body where the
distribution is roughly uniform, and the excess of counts in the
small-$p$-value tail, more than one standard deviation above the mean,
provides an estimate of the number of languages that can be
confidently said to violate the random-sampling model.  

From the upper panel of Fig.~\ref{fig:p_dists_marg_sat}, with a total
sample of 81 languages, we can estimate a number of $\sim 0.05 \times
81 \approx 4-5$ excess languages at the lowest $p$-values of 0.05 and
0.1, with an additional 2--3 languages rejected by the product model
in the range $p$-value $\sim 0.2$.  Comparable plots in
Fig.~\ref{fig:p_dists_marg_sat} (lower panel) for the 27 three-language
aggregate distributions are marginally consistent with random sampling
for the saturation model, as expected from
Fig.~\ref{fig:sample_KL_hist_data_27_joint} above.  We will show in
the next section that a more systematic trend in language fluctuations
with size provides evidence that the cause for these rejections is
excess variance due to repeated attachment of links to a subset of
nodes.

{\bf Excess fluctuations in degree of polysemy}

\begin{figure}[ht]
  \begin{center} 
  \includegraphics[scale=0.45]{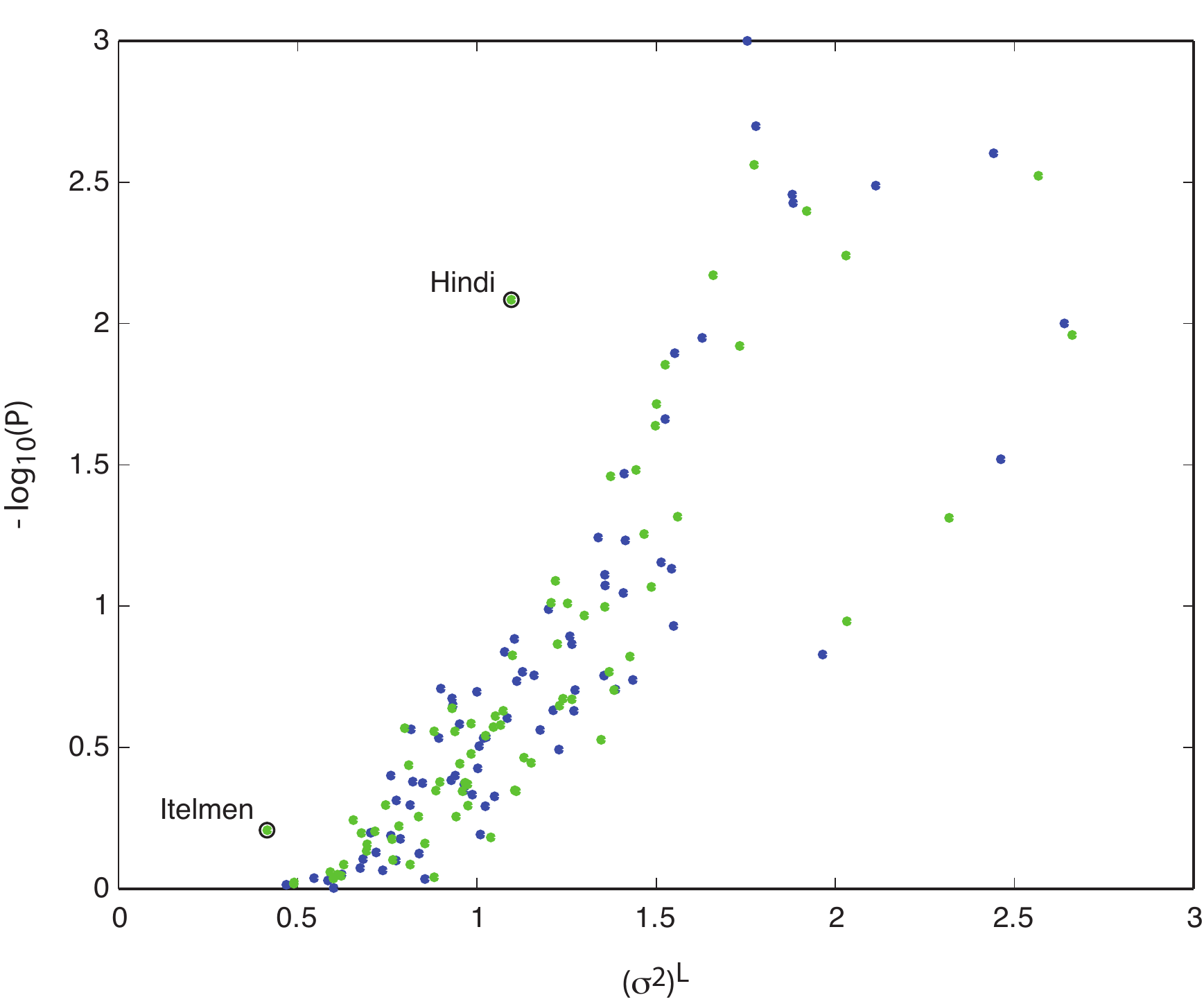} \\
  \includegraphics[scale=0.45]{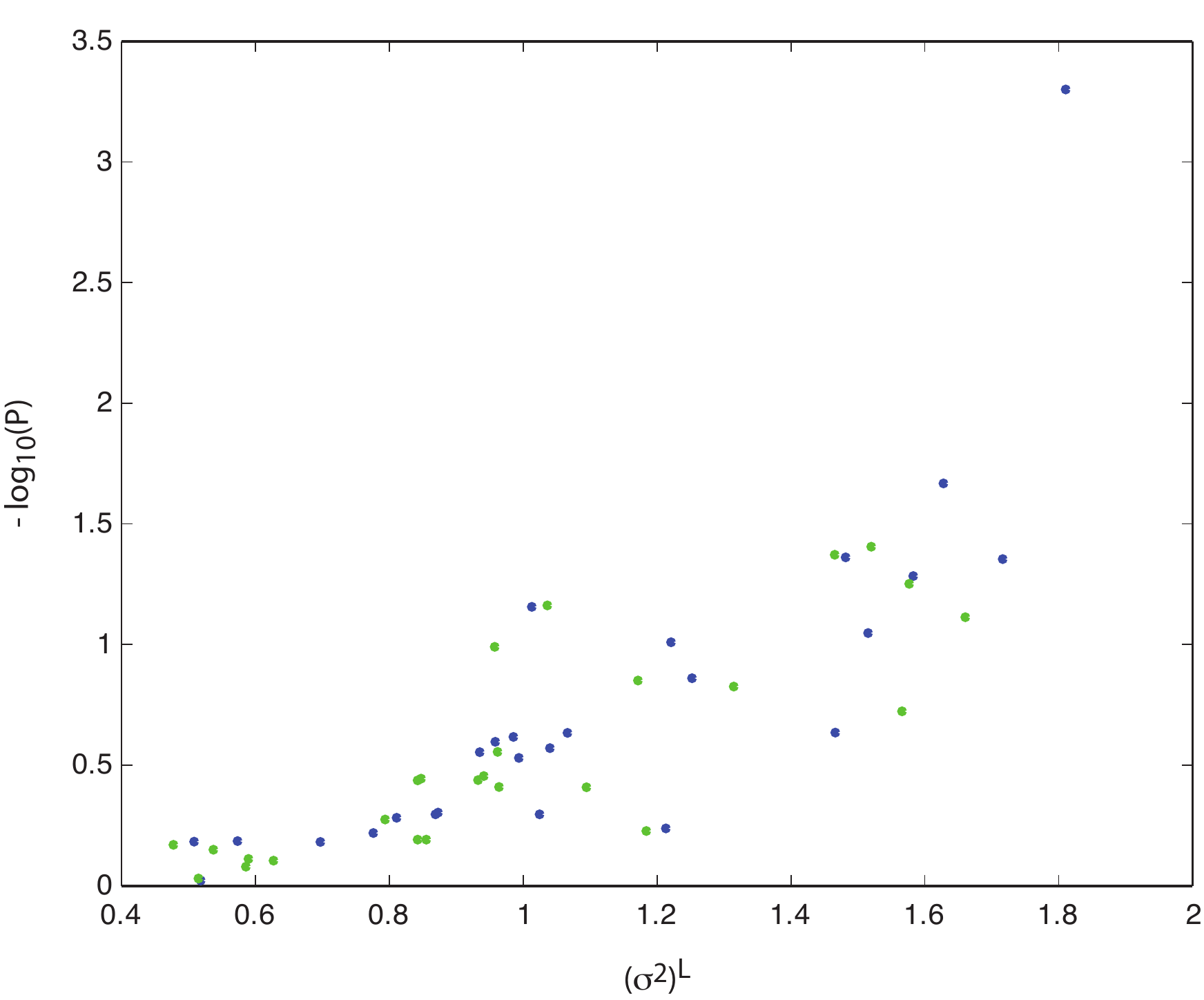}
  \caption{
    (Upper panel:)
    $-\log_{10} \! \left( P \right)$ plotted versus relative variance
    ${ \left( {\sigma}^2 \right) }^L$ from Eq.~(\ref{eq:rel_var}) for
    the 78 languages with non-zero $p$-values from
    Fig.~\ref{fig:one_lang_p_vals}.  
    (blue) saturation model; (green) product model.  
    Two languages (circled) which appear as outliers with
    anomalously small relative variance in the product model -- 
    Itelman and Hindi -- disappear into the central tendency with the
    saturation model.  (Lower panel:) an equivalent plot for 26
    three-language bins.  Notably, the apparent separation of
    individual large-$n^L$ langauges into two groups has vanished
    under binning, and a unimodal and smooth dependence of $-\log_{10}
    \! \left( P \right)$ on ${ \left( {\sigma}^2 \right) }^L$ is
    seen. 
    \label{fig:one_lang_pval_relvar_corr} 
  }
  \end{center}
\end{figure}

If we define the size-weighted relative variance of a language
analogously to the error term in Eq.~(\ref{eq:error_sat}), as 
\begin{equation}
  {
    \left( {\sigma}^2 \right)
  }^L \equiv 
  \frac{1}{n^L}
  \sum_S
  {
    \left( 
      n_S^L -
      n^L p_{S \mid L}^{\mbox{\scriptsize model}}
    \right) 
  }^2 ,  
\label{eq:rel_var}
\end{equation}
Fig.~\ref{fig:one_lang_pval_relvar_corr} shows that $-\log_{10} \!
\left( p{\rm -value} \right)$ has high rank correlation with ${ \left( {\sigma}^2
  \right) }^L$ and a roughly linear regression over most of the
range.\footnote{Recall from Eq.~(\ref{eq:D_KL_var_L}) that the leading
  quadratic term in the KL-divergence differs from ${ \left(
      {\sigma}^2 \right) }^L$ in that it presumes Poisson fluctuation
  with variance $n^L p_{S \mid L}^{\mbox{\scriptsize model}}$ at the
  level of each {\it word}, rather than uniform variance $\propto
  n^L$ across all words in a language.  The relative variance is thus
  a less specific error measure.} Two languages (Itelmen and Hindi),
which appear as large outliers relative to the product model, are
within the main dispersion in the saturation model, showing that their
discrepency is corrected in the mean link number.  We may therefore
understand a large fraction of the improbability of languages as
resulting from excess fluctuations of their degree numbers relative to
the expectation from Poisson sampling.

\begin{figure}[ht]
  \begin{center} 
  \includegraphics[scale=0.5]{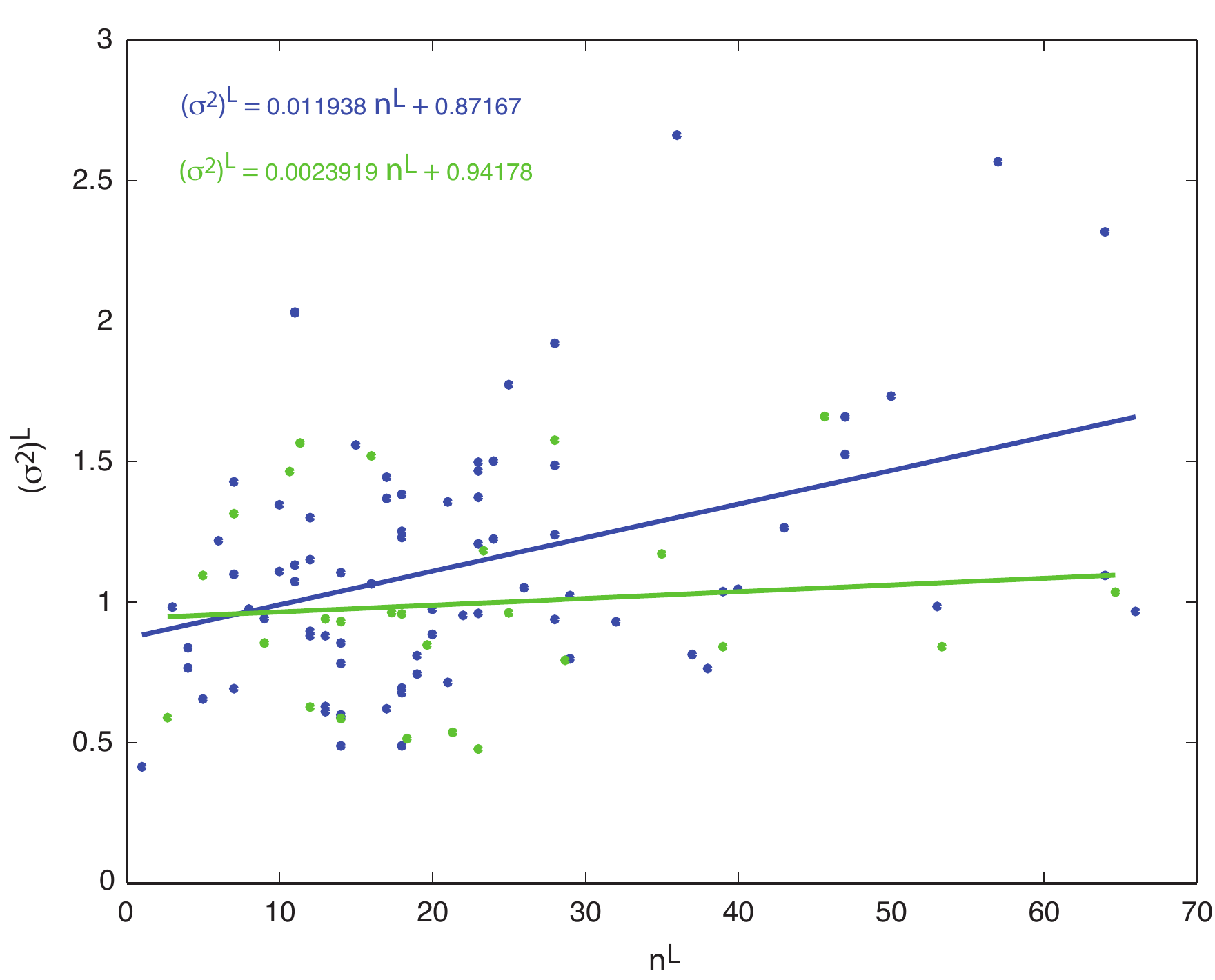}
  \caption{
    Relative variance from the saturation model versus total link
    number $n^L$ for 78 languages excluding Basque, Haida, and
    Yor\`ub\'a.  Least-squares regression are shown for three-language
    bins (green) and individual languages (blue), with regression
    coefficients inset.  Three-language bins are consistent with
    Poisson sampling at all $n^L$, whereas single languages show
    systematic increase of relative variance with increasing $n^L$.  
    \label{fig:variance_test_sat_p_value_excludes} 
  }
  \end{center}
\end{figure}

Fig.~\ref{fig:variance_test_sat_p_value_excludes} then shows the
relative variance from the saturation model, plotted versus total
average link number for both individual languages and three-language
bins.  The binned languages show no significant regression of relative
variance away from the value unity for Poisson sampling, whereas
single languages show a systematic trend toward larger variance in
larger languages, a pattern that we will show is consistent with
``clumpy'' sampling of a subset of nodes.  The disappearance of this
clumping in binned distributions shows that the clumps are
uncorrelated among languages at similar $n^L$.  

\clearpage

{\bf Correlated link assignments}

We may retain the mean degree distributions, while introducing a
systematic trend of relative variance with $n^L$, by modifying our
sampling model away from strict Poisson sampling to introduce
``clumps'' of links.  To remain within the use of minimal models, we
modify the sampling procedure by a single parameter which is
independent of word $S$, language-size $n^L$, or particular language
$L$.

We introduce the sampling model as a function of two parameters, and
show that one function of these is constrained by the regression of
excess variance. (The other may take any interior value, so we have an
equivalence class of models.)  In each language, select a number
$\mathcal{B}$ of Swadesh entries randomly.  Let the Swadesh indices be
denoted ${\left\{ S_\beta \right\}}_{\beta \in 1 , \ldots \mathcal{B}}$.  
We will take some fraction of the total links in that language, and 
assign them only to the Swadeshes whose indices are in this privileged 
set.  Introduce a parameter $q$ that will determine that fraction.

We require correlated link assignments be consistent with the mean 
determined by our model fit, since binning of data has shown no
systematic effect on mean parameters.  Therefore, for the random
choice ${\left\{ S_\beta \right\}}_{\beta \in 1 , \ldots
\mathcal{B}}$, introduce the normalized density on the privileged
links 
\begin{equation}
  {\pi}_{S \mid L} \equiv 
  \frac{
    p_{S \mid L}^{\mbox{\scriptsize model}}
  }{
    \sum_{\beta = 1}^{\mathcal{B}}
    p_{S_{\beta} \mid L}^{\mbox{\scriptsize model}}
  }
\label{eq:pi_def}
\end{equation}
if $S \in {\left\{ S_\beta \right\}}_{\beta \in 1 , \ldots
\mathcal{B}}$ and ${\pi}_{S \mid L} = 0$ otherwise.  Denote the
aggregated weight of the links in the priviledged set by 
\begin{equation}
  W \equiv 
  \sum_{\beta = 1}^{\mathcal{B}}
  p_{S_{\beta} \mid L} . 
\label{eq:W_def}
\end{equation}
Then introduce a modified probability distribution based on the
randomly selected links, in the form
\begin{equation}
  {\tilde{p}}_{S \mid L} \equiv  
  \left( 1 - q W \right)
  p_{S \mid L} + 
  q W 
  {\pi}_{S \mid L} . 
\label{eq:p_tilde_def}
\end{equation}
Multinomial sampling of $n^L$ links from the distribution
${\tilde{p}}_{S \mid L}$ will produce a size-dependent variance of the
kind we see in the data.  The expectated degrees given any particular
set ${\left\{ S_\beta \right\}}$ will not agree with the means in the
aggregate graph, but the ensemble mean over random samples of
languages will equal $p_{S \mid L}$, and binned groups of languages
will converge toward it according to the central-limit theorem. 

The proof that the relative variance increases linearly in $n^L$ comes
from the expansion of the expectation of Eq.~(\ref{eq:rel_var}) for
random samples, denoted 
\begin{eqnarray}
\lefteqn{
  \left< 
    {
      \left( 
        {\hat{\sigma}}^2 
      \right)
    }^L
  \right> \equiv 
  \left< 
    \frac{1}{n^L}
    \sum_S 
    {
      \left( 
        {\hat{n}}_S^L - 
        n^L
        p_{S \mid L}^{\mbox{\scriptsize model}} 
      \right)
    }^2  
  \right>
} & & 
\nonumber \\
& = & 
  \left< 
    \frac{1}{n^L}
    \sum_S 
    {
      \left[ 
        \left( 
          {\hat{n}}_S^L - 
          n^L
          {\tilde{p}}_{S \mid L} 
        \right) + 
        n^L 
        \left( 
          {\tilde{p}}_{S \mid L} - 
          p_{S \mid L}^{\mbox{\scriptsize model}} 
        \right)
      \right] 
    }^2 
  \right>
\nonumber \\
& = & 
  \left< 
    \frac{1}{n^L}
    \sum_S 
    {
      \left( 
        {\hat{n}}_S^L - 
        n^L
        {\tilde{p}}_{S \mid L} 
      \right)
    }^2 
  \right> + \nonumber\\
&&\qquad
  n^L
  \left< 
    \sum_S 
    {
      \left( 
        {\tilde{p}}_{S \mid L} - 
        p_{S \mid L}^{\mbox{\scriptsize model}} 
      \right)
    }^2 
  \right> . 
\label{eq:rel_var_clumpy}
\end{eqnarray}

The first expectation over ${\hat{n}}_S^L$ is constant (of order
unity) for Poisson samples, and the second expectation (over the sets
${\left\{ S_\beta \right\}}$ that generate ${\tilde{p}}_{S \mid L}$)
does not depend on $n^L$ except in the prefactor.  Cross-terms vanish
because link samples are not correlated with samples of ${\left\{
    S_\beta \right\}}$.  Both terms in the third line of
Eq.~(\ref{eq:rel_var_clumpy}) scale under binning as ${\left(
    \mbox{bin-size} \right)}^0$.  The first term is invariant due to
Poisson sampling, while in the second term, the central-limit theorem
reduction of the variance in samples over ${\tilde{p}}_{S \mid L}$
cancels growth in the prefactor $n^L$ due to aggregation.

Because the linear term in Eq.~(\ref{eq:rel_var_clumpy}) does not
systematically change under binning, we interpret the vanishing of the
regression for three-language bins in
Fig.~\ref{fig:variance_test_sat_p_value_excludes} as a consequence of
fitting of the mean value to binned data as sample
estimators.\footnote{We have verified this by generating random
  samples from the model~(\ref{eq:rel_var_clumpy}), fitting a
  saturation model to binned sample configurations using the same
  algorithms as we applied to our data, and then performing
  regressions equivalent to those in
  Fig.~\ref{fig:variance_test_sat_p_value_excludes}.  In about $1/3$
  of cases the fitted model showed regression coefficients consistent
  with zero for three-language bins.  The typical behavior when such
  models were fit to random sample data was that the three-bin
  regression coefficient decreased from the single-language regression
  by $\sim 1/3$.}  Therefore, we require to choose parameters
$\mathcal{B}$ and $q$ so that regression coefficients in the data are
typical in the model of clumpy sampling, while regressions including
zero have non-vanishing weight in models of three-bin aggregations.

\begin{figure}[ht]
  \begin{center} 
  \includegraphics[scale=0.5]{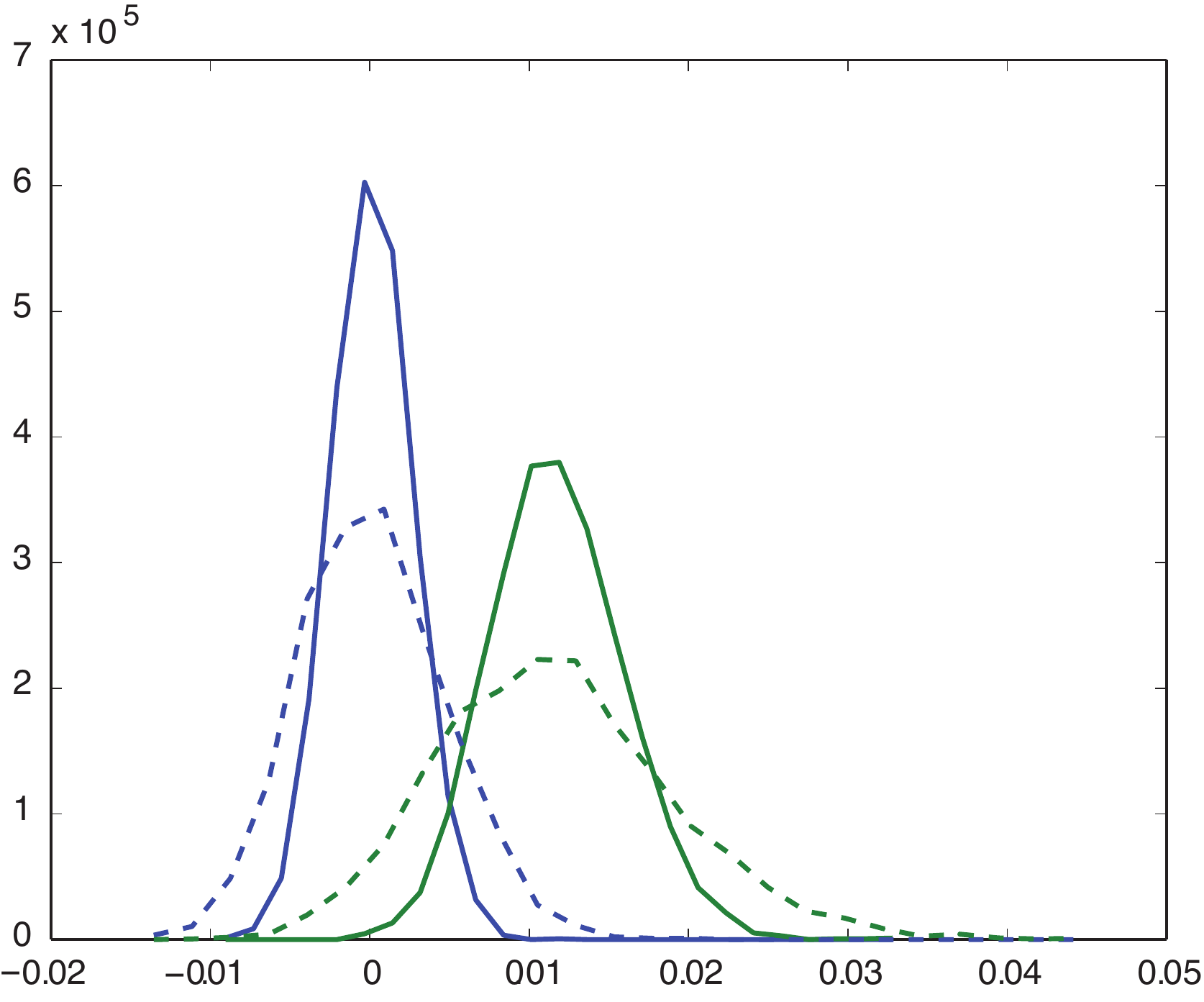}
  \caption{
    Histograms of regression coefficients for language link samples
    $\left\{ {\hat{n}}_S^L \right\}$ either generated by Poisson
    sampling from the saturation model $p_{S \mid
      L}^{\mbox{\scriptsize model}}$ fitted to the data (blue), or
    drawn from clumped probabilities ${\tilde{p}}_{S \mid L}$ defined
    in Eq.~(\ref{eq:p_tilde_def}), with the set of privileged words
    $\left\{ S_{\beta} \right\}$ independently drawn for each language
    (green).  Solid lines refer to joint configurations of 78
    individual languages with the $n^L$ values in
    Fig.~\ref{fig:variance_test_sat_p_value_excludes}.  Dashed lines
    are 26 non-overlapping three-language bins.   
    \label{fig:multinom_samp_hists_78_26_words} 
  }
  \end{center}
\end{figure}

Fig.~\ref{fig:multinom_samp_hists_78_26_words} compares the range of
regression coefficients obtained for random samples of languages with
the values $\left\{ n^L \right\}$ in our data, from either the
original saturation model $p_{S \mid L}^{\mbox{\scriptsize sat}}$, or
the clumpy model ${\tilde{p}}_{S \mid L}$ randomly re-sampled for each
language in the joint configuration.  Parameters used were
($\mathcal{B} = 7$, $q = 0.975$).\footnote{Solutions consistent with
  the regression in the data may be found for $\mathcal{B}$ ranging
  from 3--17.  $\mathcal{B} = 7$ was chosen as an intermediate
  value, consistent with the typical numbers of nodes appearing in our
  samples by inspection.}  With these parameters, $\sim 1/3$ of links
were assigned in excess to $\sim 1/3$ of words, with the remaining
$2/3$ of links assigned according to the mean distribution. 

The important features of the graph are: 1) Binning does not change
the mean regression coefficient, verifying that
Eq.~(\ref{eq:rel_var_clumpy}) scales homogeneously as ${\left(
    \mbox{bin-size} \right)}^0$.  However, the variance for binned
data increases due to reduced number of sample points; 2) the observed
regression slope 0.012 seen in the data is far out of the support of
multinomial sampling from $p_{S \mid L}^{\mbox{\scriptsize sat}}$,
whereas with these parameters, it becomes typical under $\left\{
  {\tilde{p}}_{S \mid L} \right\}$ while still leaving significant
probability for the three-language binned regression around zero (even
without ex-post fitting).

\end{document}